\documentclass[12pt]{iopart}

\usepackage{iopams}
\usepackage{graphicx}   
\usepackage{dcolumn}    
\usepackage{bm}         
\usepackage{multirow}   

\eqnobysec  

\begin{document}

\title[Slippage and Nanorheology of Thin Liquid Polymer
Films]{Slippage and Nanorheology of Thin Liquid Polymer Films}

\author{Oliver B\"aumchen\footnote{Present address: McMaster University, Department of Physics and Astronomy, 1280 Main Steet West, L8S 4M1, Hamilton, ON, Canada}, Renate Fetzer\footnote{Present address: Karlsruhe Institute of Technology, Institute for Pulsed Power and
Microwave Technology, 76344 Eggenstein-Leopoldshafen, Germany}, Mischa Klos, Matthias Lessel, Ludovic Marquant, Hendrik H\"ahl and Karin Jacobs}

\address{Department of Experimental Physics, Saarland University, D-66041 Saarbr\"ucken, Germany}
\ead{k.jacobs@physik.uni-saarland.de}

\begin{abstract}

Thin liquid films on surfaces are part of our everyday life, they
serve e.g.\ as coatings or lubricants. The stability of a thin layer
is governed by interfacial forces, described by the
effective interface potential, and has been subject of many studies
in the last decades. In recent years, the dynamics of thin liquid
films came into focus since results on the reduction of the
glass transition temperature raised new questions on the behavior of
especially polymeric liquids in confined geometries. The new focus
was fired by theoretical models that proposed significant
implication of the boundary condition at the solid/liquid interface
on the dynamics of dewetting and the form of a liquid front. Our
study reflects these recent developments and adds new experimental
data to corroborate the theoretical models. To probe the
solid/liquid boundary condition experimentally, different ways are
possible, each bearing advantages and disadvantages, which will be
discussed. Studying liquid flow on a variety of different
substrates entails a view on the direct implications of the
substrate, the experimental focus of this study is the variation of
the polymer chain length: The results demonstrate that inter-chain
entanglements and in particular their density close to the interface,
originating from non-bulk conformations, govern liquid slip of a
polymer.

\end{abstract}



\submitto{\JPCM}


\maketitle


\section{Introduction}
\label{Intro}

Confining a material often induces new phenomena that cannot be
found in bulk. Prominent examples are the physics of a quantum dot
or the surface reconstructions of many crystalline materials.
Confinement in liquid flow geometries, however, came into focus with
the rise of the field of nano- and microfluidics and with its
application in lab-on-a-chip devices for transporting, mixing,
separating or analyzing minute amounts of liquids
\cite{Squ05,Song03,Jac05,Fra11}. Recent developments tend to avoid
huge external features such as pumps to control the flow by
designing analogues to capacitors, resistors or diodes that are
capable to control currents in electronic circuits \cite{Les09}.
Controlling the flow also involves a control of the solid/liquid
boundary condition. In recent years, various experimental methods
have been developed in order to characterize the boundary condition
to more than "no slip" and "full slip" \cite{Net05,Lau07,Boc07}. At
the same time, theoretical descriptions have advanced to include
slip effects into the classical thin film equations
\cite{Kar04,Mue05,Blo08}, which now enable a prediction of the
influence of slippage to various other situations, e.g.\ to the
Rayleigh-Plateau instability \cite{Mue052,Kin09,Mue11} or to
spinodal dewetting \cite{Rau08}. Combining the experiments with those 
theories now allows for new insights into confinement effects in
liquids, some of which will be reported in this study.

Liquid slip is characterized by the slip length $b$ and is defined
as the extrapolation length of the velocity gradient to zero
velocity \cite{Nav23}. Studies of published data of $b$ can roughly
be divided into two groups, the ones reporting on slip effects in
the tens of nm-range \cite{Pit00,Zhu02,Cot02,Cot03,Cho04,Kon10} and
the ones observing large slip effects in the micron or the 'plug
flow' range
\cite{Tre02,Fet05,Fet06,Fet07,Fet073,Bae08,Bae09,Bae092}. Especially
for high molecular weight (entangled) polymer melts, large slip
lengths were expected according to de Gennes' prediction
\cite{deG79}. Moreover, de Gennes proposes the slip length $b$ to increase proportional to the
third power of the chain length $N$ (or the molecular weight,
respectively), i.e.\ $b=aN^3/N_\mathrm{e}^2$, where $a$ denotes the
molecular size and $N_\mathrm{e}$ the entanglement length of the
polymer. For this study, we especially took care that we will be
able to probe the entire range of slippage, from no to full slip,
and designed the experiments accordingly.

\begin{figure}[b]
\begin{center}
\includegraphics[width=0.4\textwidth]{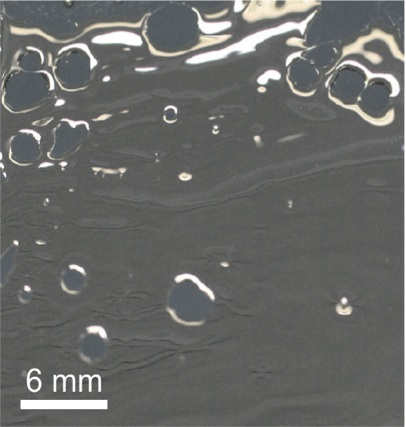}
\includegraphics[width=0.4\textwidth]{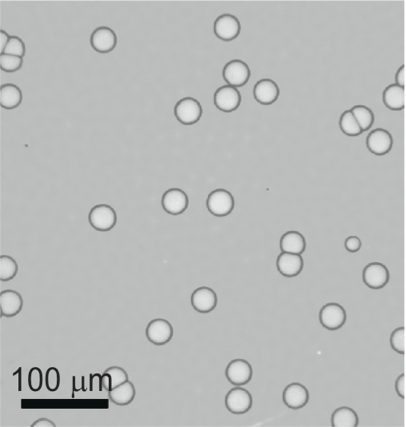}
\end{center}
\caption{Dewetting films: Honey (left) dewetting from a hydrophobic (OTS-covered Si wafer)
surface at room temperature and a thin PS(65k) film (right)
dewetting from a hydrophobized (AF\,1600 layer) Si wafer at
140$\,^{\circ}$C. A typical distribution of nucleated holes is
observed.} \label{honey-and-PS}
\end{figure}

To probe slip effects in confined fluids, thin films of polymer
melts were prepared on extremely smooth, well-controlled substrates,
typically Si wafers with various hydrophobic coatings. To illustrate
dewetting, Fig.~\ref{honey-and-PS} depicts on the left side a
snapshot of honey dewetting from a hydrophobic surface and on the
right side the lab situation, a thin liquid polystyrene film
dewetting from a hydrophobized Si wafer. For both situations, mass
conservation is valid and all liquid that once was inside the holes
is now accumulated in a rim around the holes. As time proceeds, the
holes grow. If two holes meet, they share a common rim (visible as a
straight ridge) that later decays into a row of droplets via a
Rayleigh-Plateau instability. The stability conditions and the
dewetting pathways can be characterized according to standard
methods \cite{Vri66,Her98,See012,See013,Jac08}. Details for the
dewetting morphologies, such as nucleation and spinodal dewetting,
can be found in our recent publications \cite{Jac08,Bae102} and
references therein. For the characterization of the boundary
condition in the way it will be presented in the following, the type
of dewetting scenario is not relevant; the only precondition is that
the liquid does not wet the substrate. For all experiments shown in
the following, holes in the film were initiated by nucleation.

The aim of the paper is to present the most recent results in a
consistent way in order to sharpen the conception for slippage
phenomena. We therefore concentrate on the experimental results and
for most of the theoretical details behind the analysis, the reader
is referred to earlier studies.


\section{Preparation}
\label{Preparation}

As polymers, mostly atactic polystyrene (PS) was used (purchased
from PSS, Mainz, Germany), but in some cases also block-copolymers
are utilized as specified later (c.f.\ section~\ref{FurtherExperiments}). Atactic
PS is preferred as it does not crystallize, the water uptake is
negligible and it can be purchased with very low polydispersity and
high purity. The molecular weight $M_\mathrm{w}$ of PS ranged from
5.61~kg/mol (termed 'PS(5.61k)') to 390~kg/mol. Polymers are
spin-cast onto mica out of a good solvent, which is toluene
(Selectipur$^{\circledR}$ or LiChrosolv$^{\circledR}$ Line by Merck,
Darmstadt, Germany) for PS. The resulting film thicknesses are in the range of 100 to 140\,nm, unless otherwise noted. Aside from single experiments presented in
sections~\ref{exsitu} and \ref{sliplength}, PS films exhibiting
molecular weights $M_\mathrm{w}\geq 35.6$k were pre-annealed on mica.
For the choice of appropriate pre-annealing conditions, we made use of experiments presented in Ref.~\cite{Pod01}, where above the entanglement length of PS the density of nucleated holes was found to decrease as the annealing time increased. Sites of especially high stress might act as nuclei for the generation of holes and their relaxation is correlated to the significant reduction of the hole density. For the largest molecular weight that was used for dewetting experiments, i.e.\ PS(390k), annealing for 3\,h at 140$^{\circ}$C did not lead to a further reduction of the hole density, as a plateau value is reached. To safely exclude the presence of internal stresses, the aforementioned pre-annealing parameters (3\,h at 140$^{\circ}$C on mica) were chosen in a consistent manner prior to dewetting. The thin film was then floated onto a MilliQ\texttrademark \,water surface from where it could be picked up with the sample surface of choice.

Si wafers (Wacker Siltronic AG, Burghausen, Germany and Si-Mat Silicon Materials, Kaufering, Germany) have been
carefully cleaned using standard procedures \cite{Si_wafer} and
hydrophobized by octadecyl-trichlorosilane (OTS),
hexadecyl-trichlorosilane (HTS) or dodecyl-trichlorosilane (DTS)
self-assembled monolayers (SAMs) using standard silanization methods
\cite{Sag80,Brz94,Les11}. To achieve a hydrophobic coating which is not based
on silanes, thin Teflon$^{\circledR}$ films were prepared by spin
coating a solution of AF\,1600
(Poly[4,5-difluoro-2,2-bis(trifluoromethyl)-1,3-dioxide-\textit{co}-tetrafluoroethylene],
Aldrich) in a perfluorinated solvent (FC-75$^{\circledR}$, Acros
Organics). Heating the samples above the glass transition
temperature $T_\mathrm{g}$ of the PS film initiates dewetting.

\begin{table}[b]
\begin{center}
\caption{\label{tabsubstrateproperties}Properties of AF\,1600, OTS,
HTS and DTS substrates: thickness $d$ of the hydrophobic layer,
$rms$ roughness, advancing water contact angle
$\theta_\mathrm{adv}$, water contact angle hysteresis
$\Delta\theta$, surface energy $\gamma_\mathrm{sv}$, static
contact angle $\theta_\mathrm{Y}$ of PS and absolute value of the spreading coefficient $S$ for PS.}
\begin{tabular}{c||c|c|c|c|c|c|c}
layer & $d$ & $rms$ & $\theta_\mathrm{adv}$ &
$\Delta\theta$ & $\gamma_\mathrm{sv}$ & $\theta_\mathrm{Y}$(PS) & $|S|$(PS)\\
 & (nm) & (nm) & ($^{\circ}$) & ($^{\circ}$) & (mN/m) & ($^{\circ}$) & (mN/m)\\
\hline
AF\,1600  &  20(2) &  0.30(3) &  128(2) & 10 & 15.0 & 88(2) & 29.7(11)\\
OTS  &  2.3(2) &  0.09(1) &  116(1) & 6 & 23.9 & 67(3) & 18.8(15)\\
HTS  &  2.0(1) &  0.16(2) &  109(1) & 10 & 24.5 & 62(2) & 16.4(10)\\
DTS  &  1.5(2) &  0.13(2) &  114(1) & 5 & 26.4 & 67(3) & 18.8(15)\\
\end{tabular}
\end{center}
\end{table}

Tab.~\ref{tabsubstrateproperties} gives details of the substrate
properties: $rms$ roughness (taken from a 1\,$\mu$m$^2$ atomic force
microscopy scan), water contact angles, static PS contact angle
$\theta_\mathrm{Y}$(PS) and the surface energies
$\gamma_\mathrm{sv}$. The latter was calculated from contact angle measurements of apolar liquids. Using these parameters together with the
dielectric constants and refractive indices of all contributing
materials, the van-der-Waals (vdW) part of the effective interface
potential $\phi$ can be inferred \cite{Isr92} for each of the
hydrophobized wafers (see Ref.~\cite{Bae102}). Alternatively,
experimental access to $\phi$ can be gained recording the spinodal
wavelength as function of film thickness, as shown in previous
studies \cite{Her98,See012,See013,Jac08}. The effective interface
potential summarizes all interfacial forces and describes the energy
needed (or gained) to bring two interfaces to a certain distance. In
our case, the distance between the two interfaces is the thickness
$h$ of the liquid film. At values of $h$ where the second derivative
of $\phi$ with respect to $h$, $\phi''(h)$, is negative, spinodal
dewetting can take place \cite{Vri66,Her98,See012}.


\section{Dewetting as a Probe to Analyze Slippage and Nanorheology}
\label{Results}

\begin{figure}[b]
\begin{center}
\includegraphics[width=0.15\textwidth]{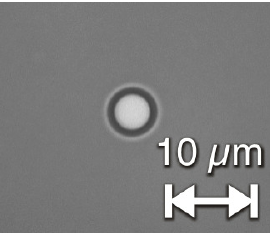}
\includegraphics[width=0.15\textwidth]{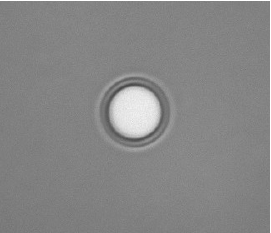}
\includegraphics[width=0.15\textwidth]{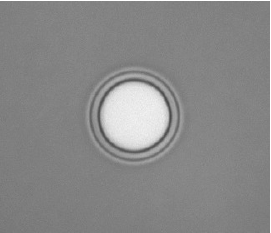}
\includegraphics[width=0.15\textwidth]{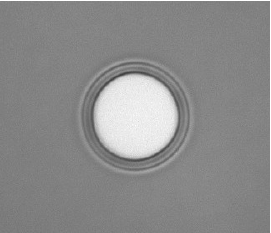}
\\
\vspace{0.1cm}
\includegraphics[width=0.15\textwidth]{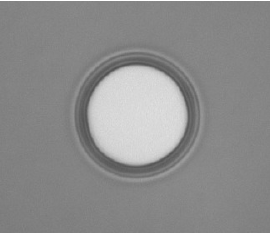}
\includegraphics[width=0.15\textwidth]{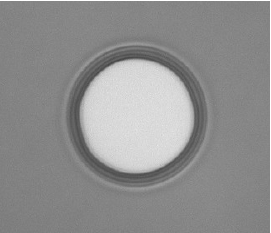}
\includegraphics[width=0.15\textwidth]{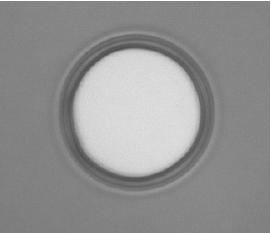}
\includegraphics[width=0.15\textwidth]{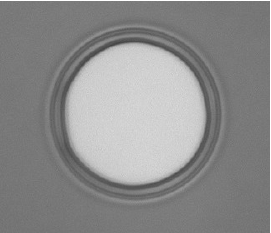}
\end{center}
\caption{Growth of a hole in a PS(13.7k) film dewetting from a
HTS-covered Si wafer at 120$\,^{\circ}$C. Time interval between
subsequent optical images (starting top left and ending bottom right): 80\,s.} \label{graphLoecheralternativ}
\end{figure}

\begin{figure}[b]
\begin{center}
\includegraphics[width=0.45\textwidth]{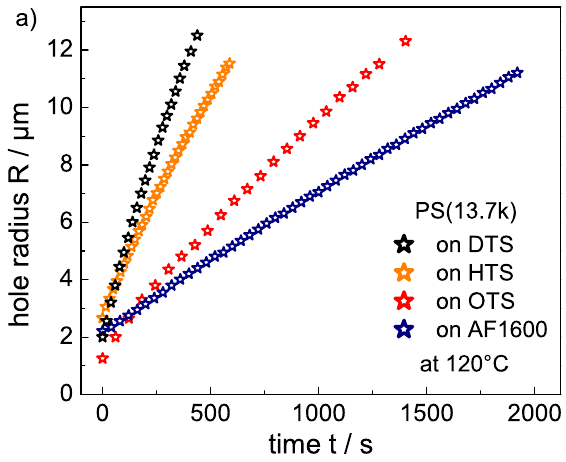}
\includegraphics[width=0.467\textwidth]{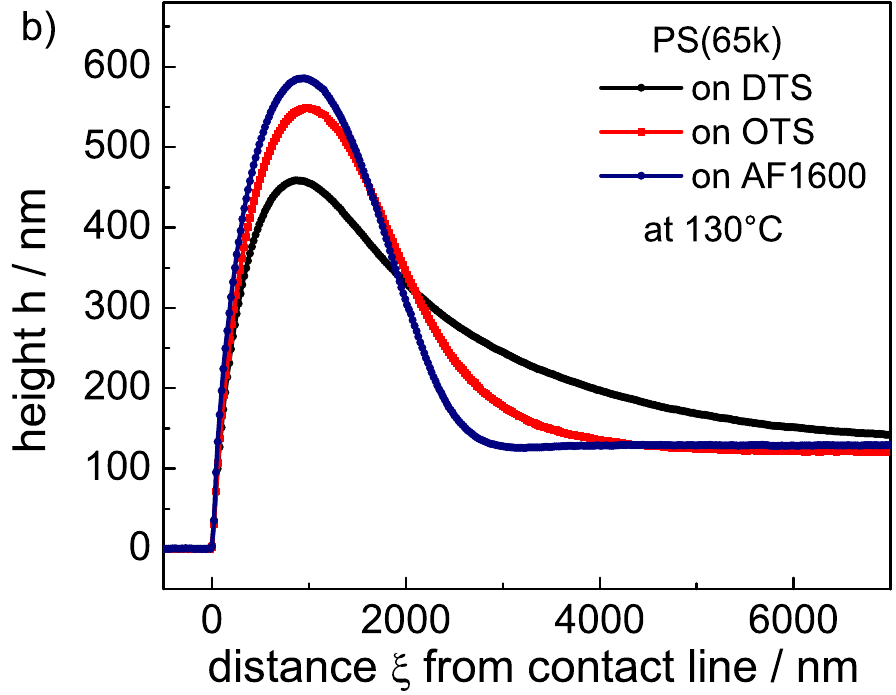}
\end{center}
\caption{a) Hole growth dynamics in a thin PS(13.7k) film on various
hydrophobic substrate surfaces (DTS, HTS, OTS and AF1600) at
120$\,^{\circ}$C. Here,  $t$=0 is defined for each experiment by the first observation of the hole. b) AFM rim profiles (c.f.\ section~\ref{exsitu}) of
thin PS(65k) films recorded at the same hole radius on DTS, OTS and
AF1600 substrates.} \label{rversust}
\end{figure}

Fig.~\ref{honey-and-PS} right depicts a typical distribution of
heterogeneously nucleated holes in a PS(65k) film, whereas Fig.~\ref{graphLoecheralternativ} shows a
temporal series of optical micrographs illustrating the growth of a
single hole. The impact of the solid/liquid interface to the
dewetting dynamics is made obvious in Fig.~\ref{rversust}a,
where the radius $R$ of a hole is recorded as function of time for
four different types of hydrophobic layers on top of a Si wafer. For the data shown, experimental parameters like PS film thickness (130\,nm), dewetting temperature
(120$\,^{\circ}$C), molecular weight (13.7~kg/mol), and roughness of
the solid surface ($rms$\,$\leq$\,0.3\,nm, c.f.\
Tab.~\ref{tabsubstrateproperties}) were held constant. Clearly,
dewetting on the DTS-covered surface proceeds the fastest. The result is surprising, since the driving force of dewetting - as characterized by the absolute value of the spreading coefficient $S$, i.e.\ $|S|=|\gamma_\mathrm{lv}(cos{\theta_\mathrm{Y}}-1)|$, is largest for AF\,1600 and lowest for HTS (c.f.\ Tab.~\ref{tabsubstrateproperties}).

Comparing identically prepared liquid films dewetting from different hydrophobic substrates highlights the fact that the flow properties and the hydrodynamic boundary condition at the solid/liquid interface are the key to understand the thin film dynamics. This entails the structural and dynamic properties of the interfacial region imposed by the substrate and, moreover, also the role of the liquid itself. The latter issue is the main subject of this study: We will focus on one hydrophobic substrate, AF1600, and systematically vary the liquid properties, in particular the length of the polymer chains as given by the molecular weight of the melt.

Dewetting experiments allow the characterization of the slip length either by analyzing the dewetting dynamics \cite{Fet073,Bae08} or by evaluating the shape of a dewetting front \cite{Fet05,Fet06,Fet07,Bae092}. As illustrated in Fig.~\ref{rversust}b, the shape of a liquid
rim (resulting from identical films and recorded at the same hole radius) is sensitive to the type of substrate and, consequently, the boundary condition \cite{Bae09}. Again, different substrates are qualitatively compared for the identical liquid (molecular weight, viscosity). A recent review describes comprehensively the two
mentioned methods to gain the slip length and gives detailed views on slip effects in thin polymer films \cite{Bae10}. The main advantage of deriving the slip length from the dynamics and the morphology of dewetting films is that the method is non-invasive: no external shear has to be applied, no colloidal probe or tracer particles interfere with the liquid.


\subsection{Characterizing the Growth of Holes}
\label{holegrowth}

\begin{figure}[b]
\begin{center}
\includegraphics[width=0.48\textwidth]{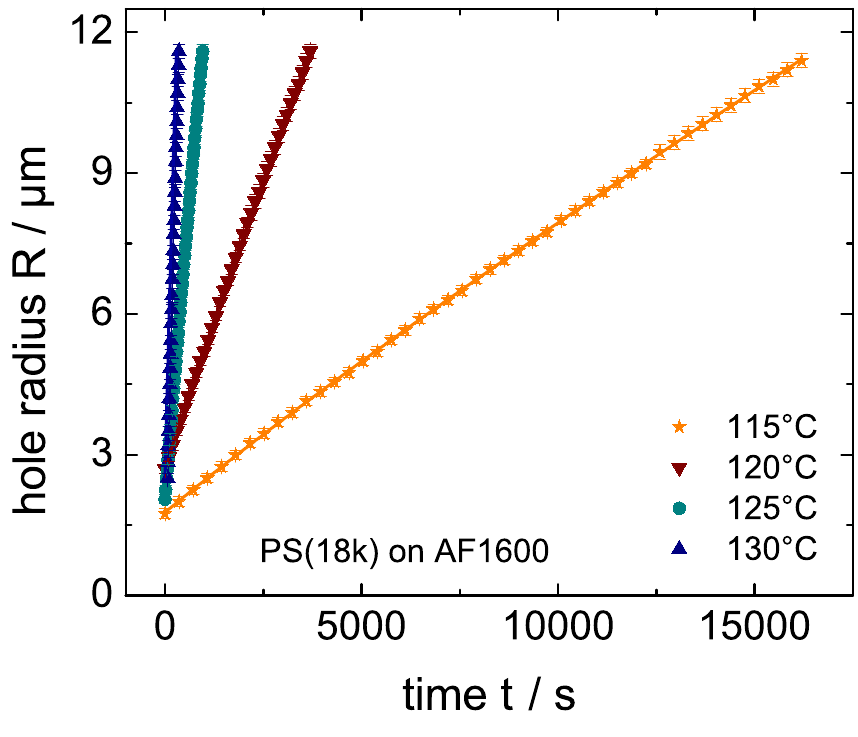}
\includegraphics[width=0.48\textwidth]{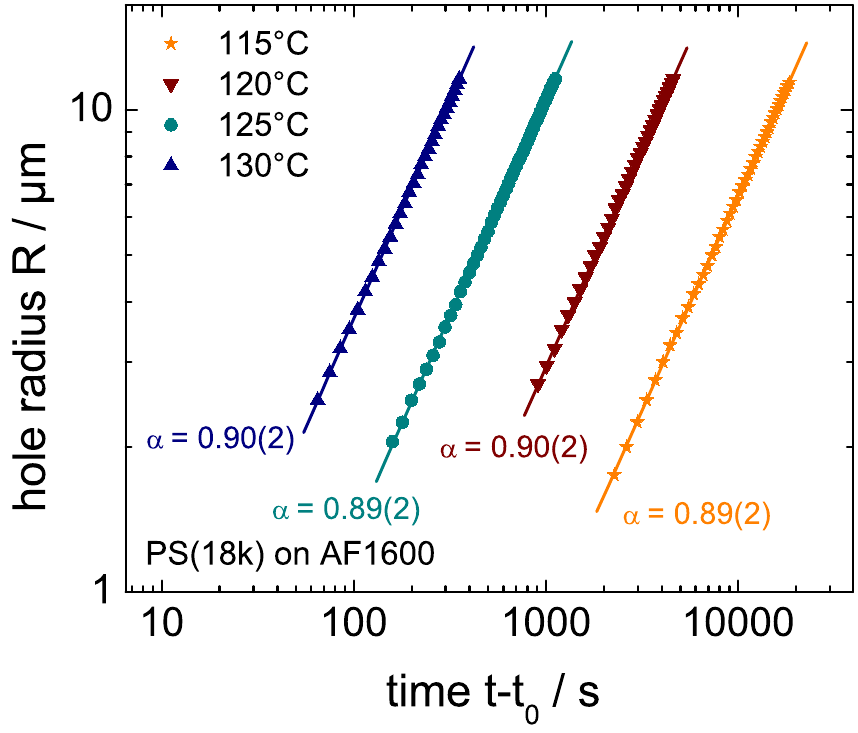}
\includegraphics[width=0.48\textwidth]{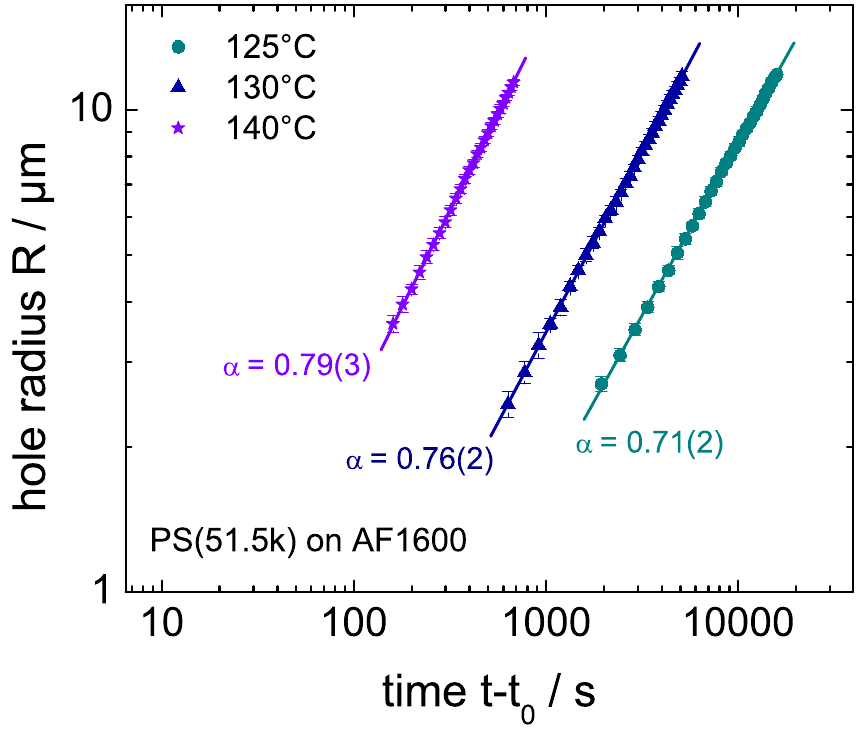}
\includegraphics[width=0.48\textwidth]{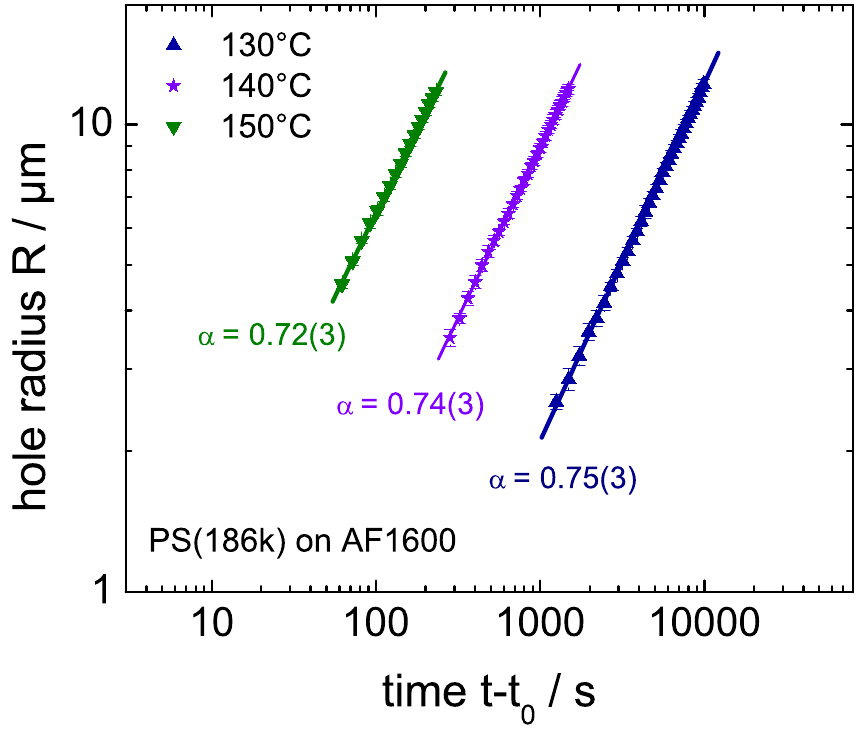}
\end{center}
\caption{Hole radius $R$ versus time $t$ in a linear (upper left) and logarithmic representations obtained from optical micrographs for PS(18k), PS(51.5k) and PS(186k) films at different dewetting temperatures. Solid lines represent fit curves of the algebraic growth function $R\propto (t-t_\mathrm{0})^{\alpha}$ to the experimental data. Note that a shift of the time scale is necessary (the condition $R(t=0)=0$ has to be matched) to obtain a linear relation in the double logarithmic representation.} \label{graphc4}
\end{figure}

From theory \cite{Red91,Red94,Bro94}, one expects a linear growth of
the hole radius $R$ with time for the no-slip situation and
$R(t)\propto t^{\alpha}$ with $\alpha=2/3$ in the case of full
slippage. Recording $R(t)$ curves in holes of dewetting polymer
films (c.f.\ Fig.~\ref{graphLoecheralternativ}), $\alpha$ is readily
determined by fitting an algebraic power law, i.e.\ $R\propto
(t-t_\mathrm{0})^{\alpha}$. Alternatively, in a double-logarithmic
plot, the $R(t-t_\mathrm{0})$ curves are fitted by a line of slope $\alpha$, c.f.\
Fig.~\ref{graphc4} \cite{mature}. The figure exemplarily illustrates
a set of typical $R(t)$ measurements in various PS films supported
by AF\,1600 substrates. Note that $t=0$ marks the time where the
hole was first detected by optical microscopy; for the evaluation of
$\alpha$, the 'real birth time' of a hole, $t_0$, is a fit parameter
that is determined for each $R(t)$ curve separately. Subsequently,
the data are evaluated according to the analytical model and
analysis technique presented in Refs.~\cite{Fet073,Bae08}, which is
based on the superposition of two dissipation mechanisms
\cite{Jac98}: viscous friction within the liquid and friction at the
solid/liquid interface (slippage).

\subsubsection{Evaluation of the Dewetting Exponent}

\begin{figure}[b]
\begin{center}
\includegraphics[width=0.46\textwidth]{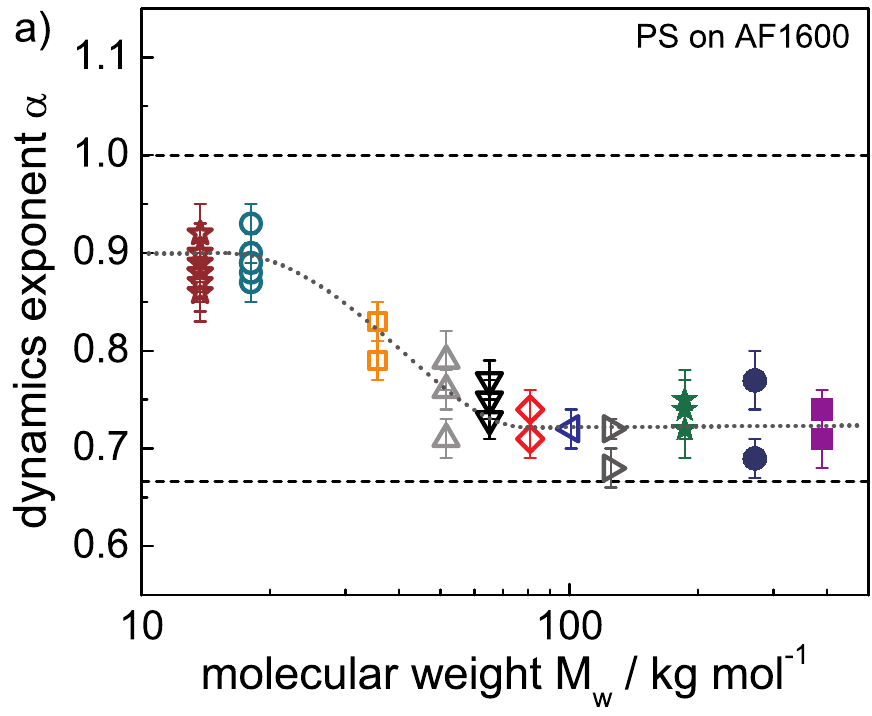}
\includegraphics[width=0.46\textwidth]{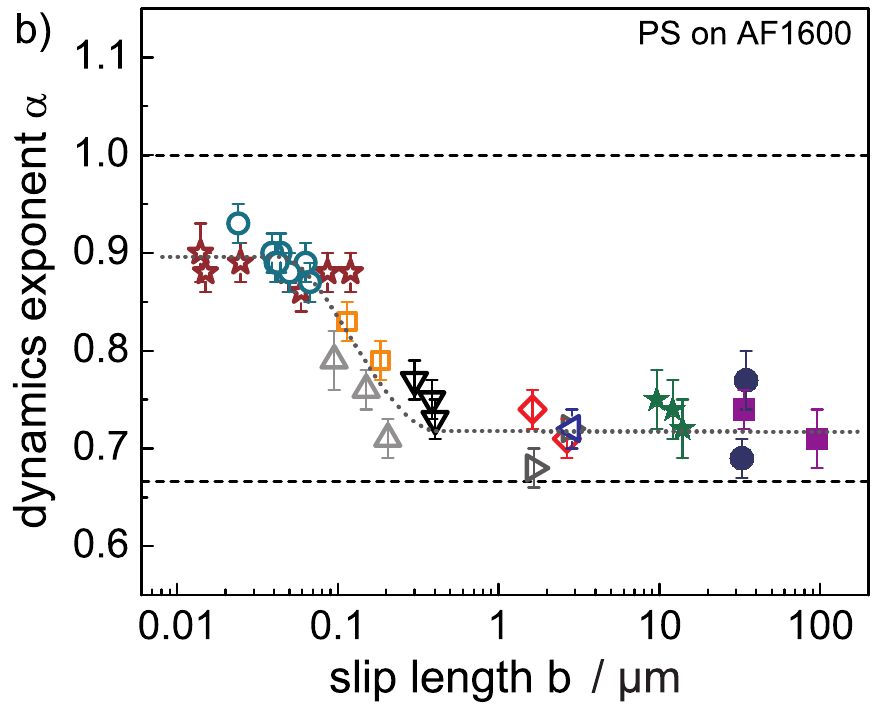}
\end{center}
\caption{a) Dewetting exponent $\alpha$ for a series of dewetting experiments using different molecular weights $M_\mathrm{w}$. b) Dewetting exponent $\alpha$ versus the slip length $b$ as obtained from the hole growth analysis (see section~\ref{holeslip}). Colors and symbols of data points for different molecular weights $M_\mathrm{w}$ in a) correspond to those used in b). Dotted lines represent guide to the eyes.} \label{graphc3}
\end{figure}

In the following, the molecular weight $M_\mathrm{w}$ of the polymer melt and the dewetting temperature were systematically varied in dewetting experiments. The values for the exponent $\alpha$ given in Fig.~\ref{graphc4} already show that $\alpha$ does not change significantly with temperature. The effect of the variation of $M_\mathrm{w}$ illustrates Fig.~\ref{graphc3}: With increasing $M_\mathrm{w}$, $\alpha$ undergoes a transition from high values around 0.9 (close to 1) to low values around 0.7 (close to 2/3). The transition occurs around the molecular weight of the polymer, where the molecules exceed the critical molecular weight $M_\mathrm{c}$ for chain entanglements. As soon as chain entanglements become considerable and slippage starts to increase, the dewetting exponent changes significantly. This finding is consistent with the theoretical expectation obtained from simple energetic considerations, which predict $R\propto t$ and $R \propto t^{2/3}$ in the limiting cases of no-slip and full-slip, respectively \cite{Red91,Red94,Bro94}. Deviations from the analytical predictions, such as  an exponent for the no-slip case being rather close to 0.9 than 1, have also been previously reported for numerical simulations \cite{Mue051}.

The experimental result, however, visualizes at the same time the limitations of this approach (c.f.\ Fig~\ref{graphc3}): Although the slip length varies by orders of magnitude, $\alpha$ does not show significant variations except for the transition from weak to strong slip (note that the slip length is orders of magnitude larger as compared to the typical film thickness). Consequently, drawing further conclusions solely from measurements of the dewetting exponent $\alpha$ with regard to slippage and the slip length in particular is not appropriate. In anticipation of the next sections, the data for $\alpha$ shown in of Fig.~\ref{graphc3}a can be related to the slip length $b$, as depicted in Fig.~\ref{graphc3}b. The exponent $\alpha$ is plotted as a function of the slip length, which was obtained via the hole growth analysis of the $R(t)$ data. The linear superposition of energy dissipation due to viscous flow and due to friction at the solid/liquid interface marks the underlying assumption for this method \cite{Jac98} and will be introduced in the following.

\subsubsection{Evaluation of the Slip Length by $R(t)$ Experiments} \label{holeslip}

Following our description as published in Refs.~\cite{Fet073,Bae08}, the contact line velocity can be expressed by two contributions, $v_v$ resulting from viscous flow and $v_\mathrm{s}$ representing liquid slip at the solid/liquid interface, given by:

\begin{equation}
\label{velocityequation}
V=v_\mathrm{v}+v_\mathrm{s}=C_\mathrm{v}(\theta)\frac{|S|}{\eta}+\frac{|S|b}{3
\eta}\frac{1}{w}
\end{equation}

and the rim width

\begin{equation}
\label{velocityequation2}
w=C_\mathrm{s}\sqrt{h_\mathrm{0}}\sqrt{R}~,
\end{equation}

where $h_\mathrm{0}$ stands for the film thickness and $C_\mathrm{v}(\theta)$ denotes the constant of proportionality characterizing the flow field in the vicinity of the three-phase contact line (receding contact angle $\theta$). $C_\mathrm{s}$ represents the constant of proportionality according to conservation of mass and can be determined via \textit{in situ} AFM measurements (c.f.\ section~\ref{insitu}), while simultaneously recording the evolution of the rim width $w$ and the radius $R$ of the hole during dewetting. By plotting the dewetting velocity $V$ (given by the first derivative $dR/dt$ of the hole radius $R(t)$) versus $1/\sqrt{R}$, the viscous and slip velocity contributions can be separated.

\begin{figure}[b]
\begin{center}
\includegraphics[width=0.5\textwidth]{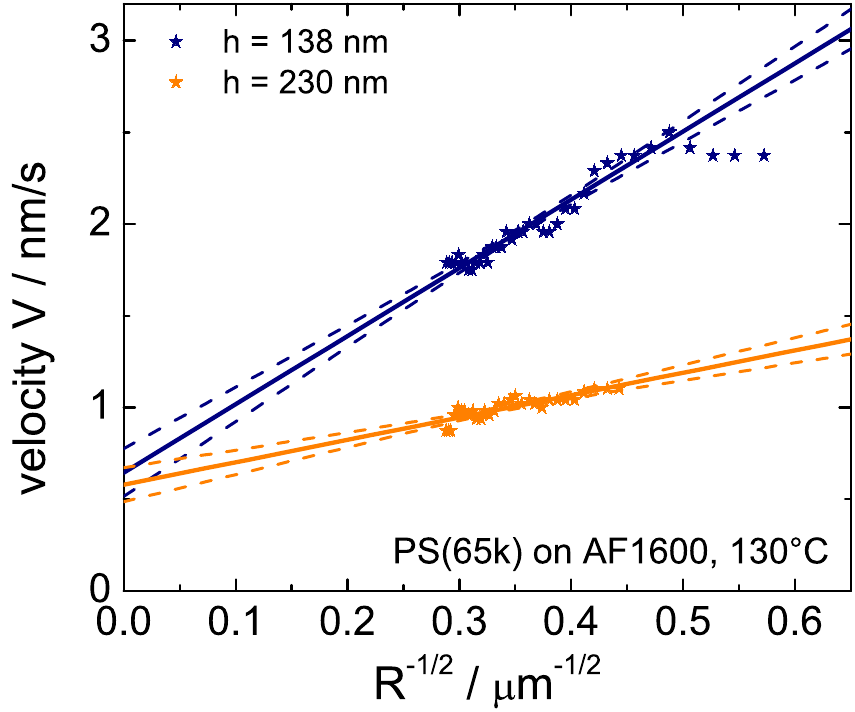}
\end{center}
\caption{Dewetting velocity plotted versus the reciprocal square root of the corresponding hole radius, i.e.\ $V(1/\sqrt{R})$. Experiments of PS(65k) dewetting from AF\,1600 for different film thicknesses at the same dewetting temperature of 130$\,^{\circ}$C and, consequently, identical viscosities. Different slopes of the linear fit curves (solid lines) can be found, the y-axis intercept $v_\mathrm{0}$, however, is identical within the experimental error for all experiments. The dashed lines illustrate the corresponding 95\% confidence bands of the fit curves.} 
\label{graphm}
\end{figure}

Fig.~\ref{graphm} exemplarily illustrates two typical $V(1/\sqrt{R})$ curves for PS(65k) films exhibiting different film thicknesses on AF\,1600 substrates. The film thickness impacts the slope of the curves, but does not act on the y-axis intercept $v_\mathrm{0}$. For small hole radii, i.e.\ for large $1/\sqrt{R}$, there are data points deviating from the linear regime. This deviation can be attributed to the presence of the pre-mature stage of hole growth, where the rim of the hole is not fully developed and therefore self-similarity of the rim profile, a precondition for the validity of the theoretical description, is not given \cite{Fet073, Bro97}.

From the slope of the $V(1/\sqrt{R})$ data representation, the slip length $b$ can be calculated via Eq.~(\ref{velocityequation}). Data for the viscosity $\eta$ were extracted from independent viscosimeter measurements \cite{seemann} by extrapolating these data \cite{WLF} according to the Williams-Landel-Ferry (WLF) equation \cite{Wil55,Rub03}. $C_\mathrm{s}$ is calculated from AFM measurements of the rim width $w$ (see section~\ref{rimprofileanalysis}) and the film thickness $h_\mathrm{0}$ in combination with an optical measurement of the hole radius $R$ after quenching the sample to room temperature \cite{Csfootnote}.

\begin{figure}[b]
\begin{center}
\includegraphics[width=0.5\textwidth]{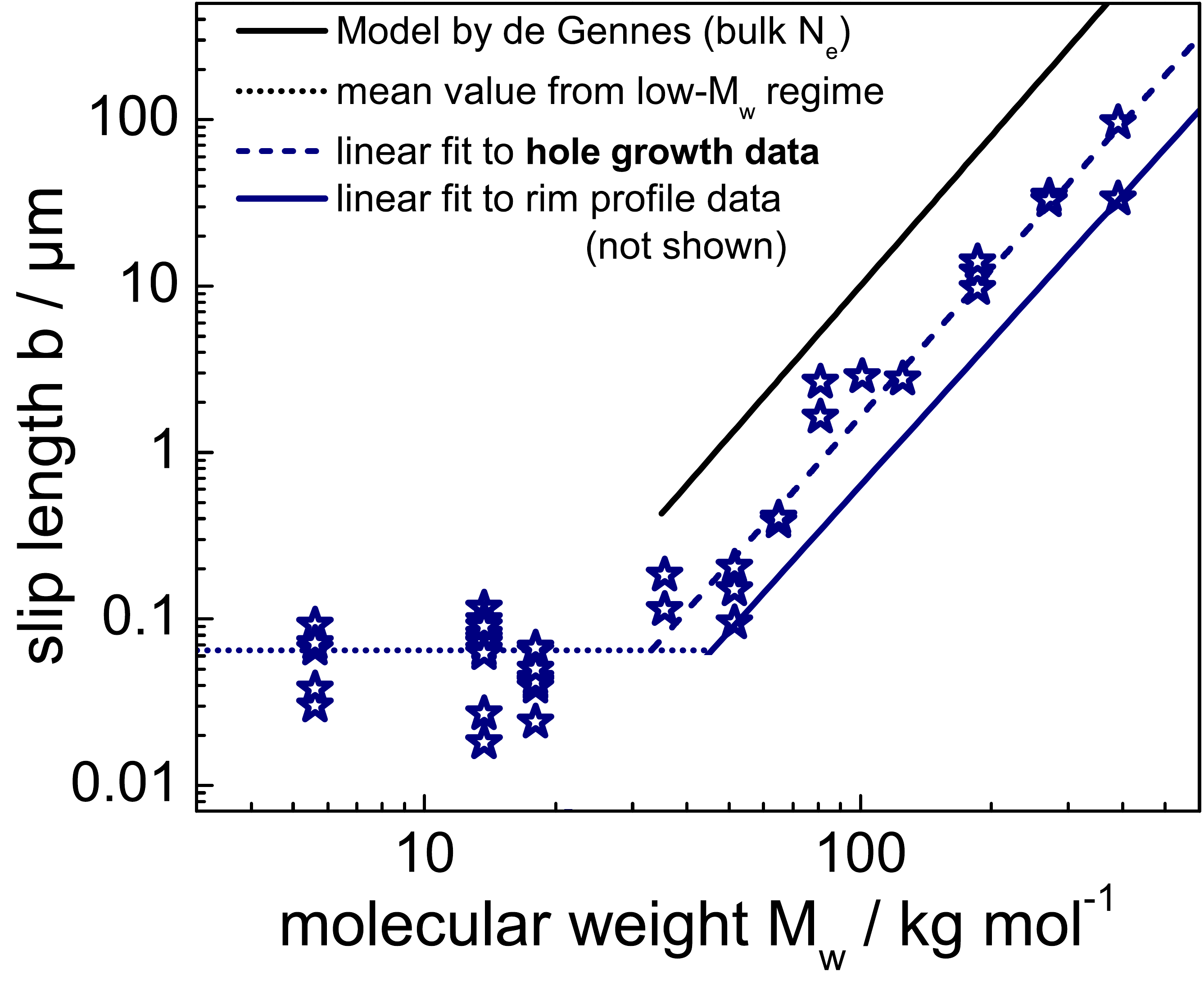}
\end{center}
\caption{Double-logarithmic plot of the slip length $b$ as function of the molecular weight $M_\mathrm{w}$ of the PS melt; $b$ is obtained from hole growth analysis for single experiments (stars) with PS dewetting from AF\,1600 substrates. The solid black line represents the model by de Gennes \cite{deG79} using bulk $N_\mathrm{e}$=163, whereas the dashed line is a fit to the experimental data. For reasons of comparison, the solid blue line represents the fit to the slip length data (not shown) obtained from the rim profile analysis (see section~\ref{sliplength}).}
\label{graphi}
\end{figure}

Fig.~\ref{graphi} illustrates the impact of the molecular weight of the polymer melt on the slip length $b$ for PS on AF\,1600 substrates: Below the critical molecular weight $M_\mathrm{c}$ for chain entanglements, slip lengths between 10 and 100\,nm are found. Above $M_\mathrm{c}$, the slip length increases with $M_\mathrm{w}^{\beta}$ and $\beta=2.9$. The values for $b(M_\mathrm{w})$, however, are systematically lower compared to the expectation by de Gennes (i.e.\ $b=aN^3/N_\mathrm{e}^2$, see Ref.~\cite{deG79}) using bulk literature values, namely $a=3{\AA}$ \cite{Red94} and $N_\mathrm{e}=163$ \cite{Rub03}. As will be discussed in section~\ref{sliplength}, these results are in good agreement with the corresponding data for $b$ obtained with the help of the rim profiles of the same holes.

\subsubsection{Evaluation of the Viscous Velocity Contribution in $R(t)$ Experiments}

\begin{figure}[b]
\begin{center}
\includegraphics[width=0.5\textwidth]{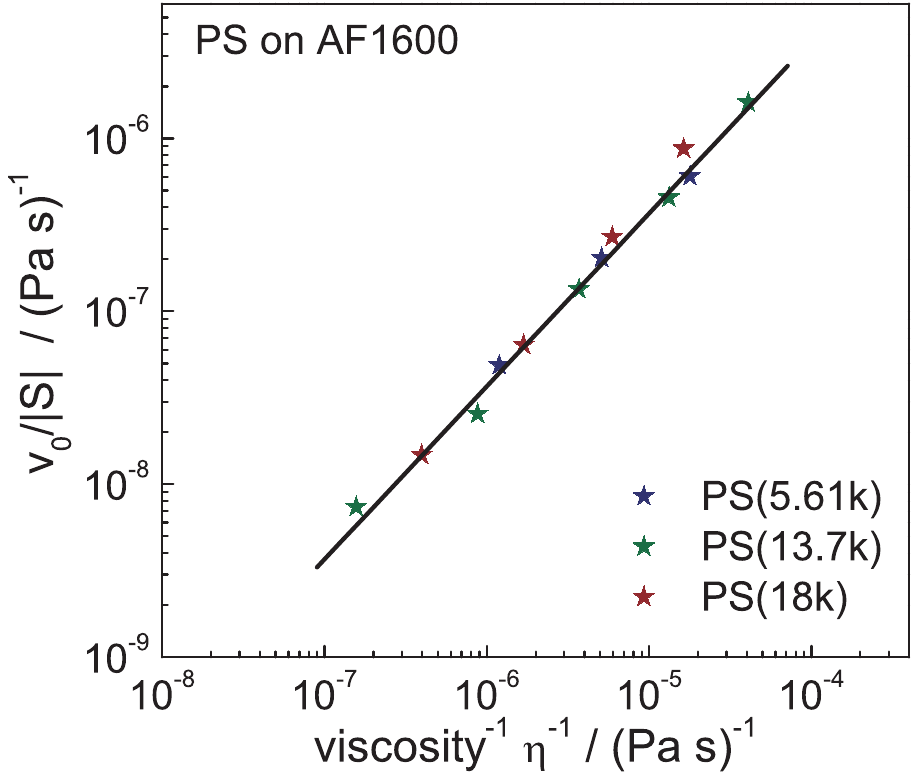}
\end{center}
\caption{Double-logarithmic representation of the y-axis intercept
$v_\mathrm{0}$, normalized by the absolute value of the spreading
coefficient $|S|$, versus the reciprocal viscosity $1/\eta$ (taken
from the solid curves in Fig.~\ref{graphvisco}). Mean values of
different experiments are plotted for different molecular weights
$M_\mathrm{w}$ below the critical molecular weight $M_\mathrm{c}$.}
\label{graphf}
\end{figure}

The position of the y-axis intercept of the extrapolated
linear fit to the experimental $V(1/\sqrt{R})$ data, c.f.\
Fig.~\ref{graphm}, is called $v_\mathrm{0}$. By plotting
$v_\mathrm{0}/|S|$ versus $1/\eta$ for different
$M_\mathrm{w}<M_\mathrm{c}$, the
predicted linear relationship between the viscous velocity
contribution $v_\mathrm{v}=v_\mathrm{0}$ and the reciprocal
viscosity $1/\eta$ is reproduced. This is illustrated in Fig.~\ref{graphf} and represents an important consistency check for the applied model. Consequently, the constant of
proportionality $C_\mathrm{v}(\theta)$ can be calculated from this
master curve for unentangled PS dewetting from AF\,1600 substrates and is determined to be $C_\mathrm{v}(\theta)=0.039$. Above entanglement, slippage changes by orders of magnitude, which has an impact on the flow field close to the contact line (c.f.\ Refs.~\cite{Red91,Red94}). Thus, although still independent of the rim size, $C_v$ does depend on the molecular weight for $M_w>M_c$.


\subsection{Characterizing the Rim Profile}
\label{rimprofileanalysis}

A rim is theoretically described by a disturbance of a liquid, that decays
exponentially into the prepared film thickness. Applying thin film
models \cite{Mue05,Fet07}, the analysis of the decay allows for a
determination of the slip length, as we have shown in earlier
studies \cite{Fet05,Fet06,Fet07,Bae09}. The rim profiles were
recorded by AFM \cite{AFM-damage}. The studies shown here focus on
PS melts of various molecular weights on top of AF\,1600. The first
part of the present section covers the investigation of the temporal
and spatial evolution of the shape of a liquid front by \textit{in
situ} AFM measurements. The second part deals with the evaluation
and interpretation of \textit{ex situ} liquid rim profiles captured
at a fixed hole radius. The latter experiments and their analysis in view of the slip length have been, in part, published in Ref. \cite{Bae092} and are extended as well as discussed in detail in the following sections.

\subsubsection{\textit{In situ} Measurement of the Evolution of the
Rim Shape} \label{insitu}

\begin{figure}[b]
\begin{center}
\includegraphics[width=0.47\textwidth]{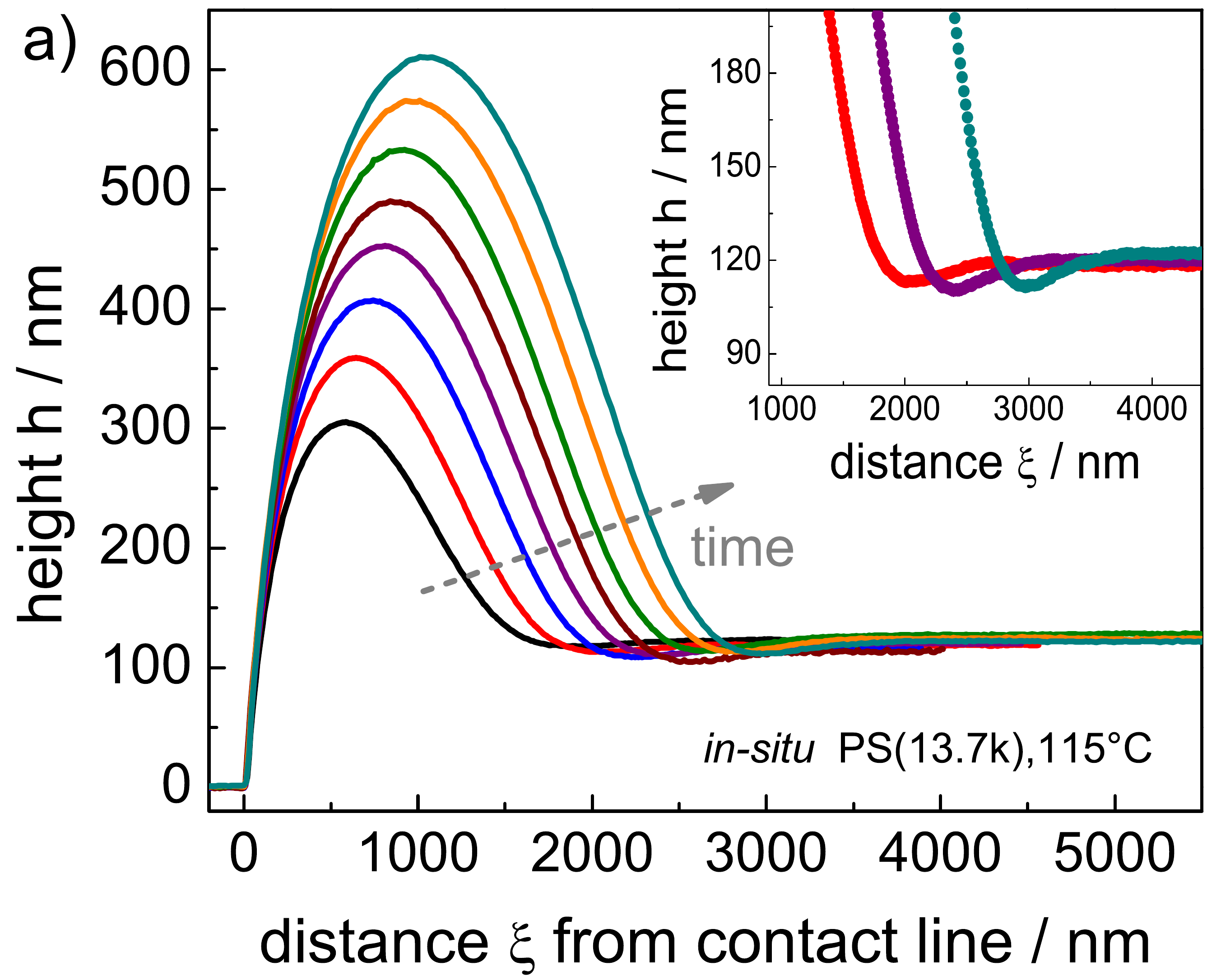}
\includegraphics[width=0.485\textwidth]{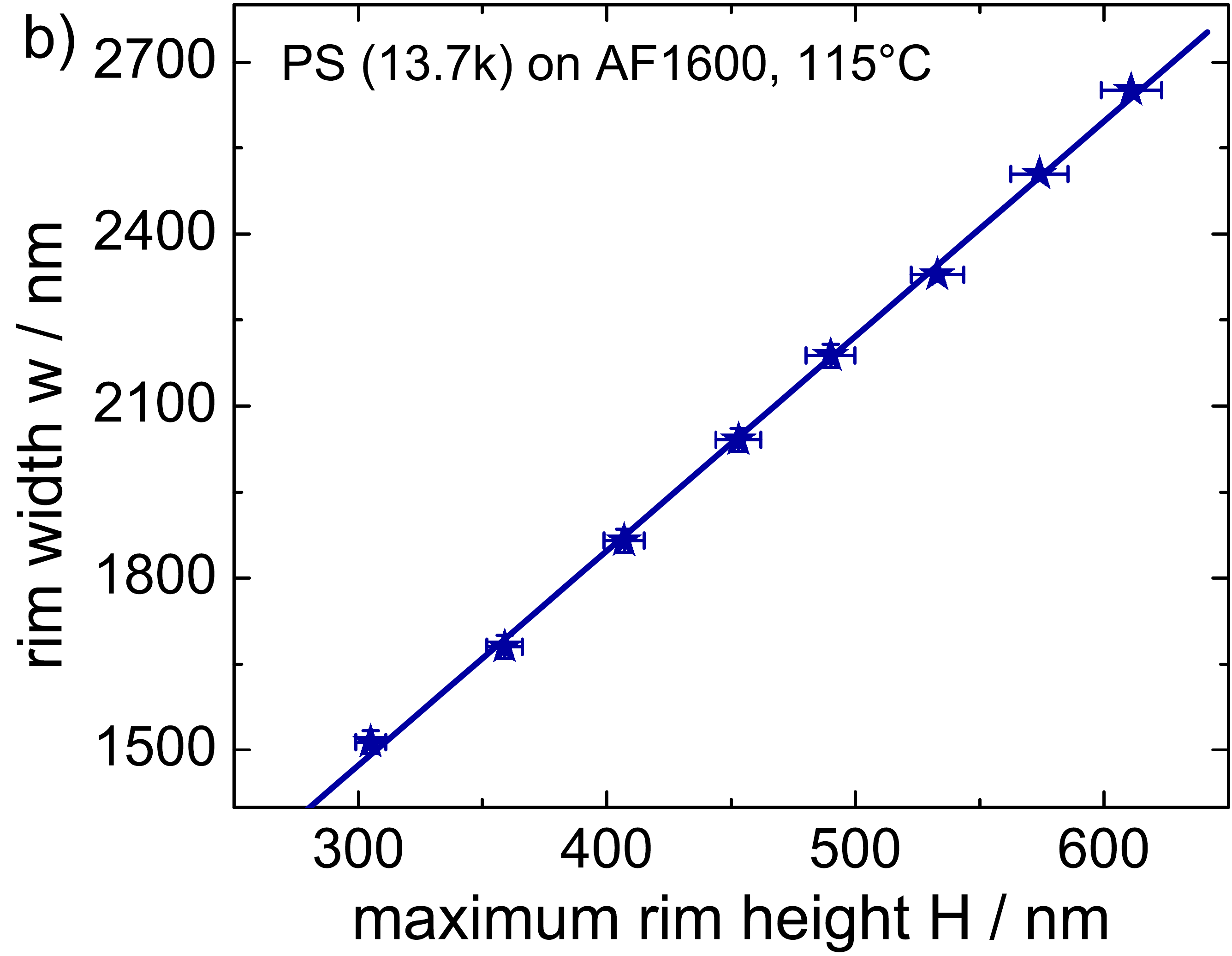}
\end{center}
\caption{a) \textit{In situ} measurement of the temporal evolution
($\Delta t=640\,s$) of the rim growth of a 120\,nm PS(13.7k) film at
115$\,^{\circ}$C on AF\,1600. All profiles clearly exhibit an
oscillatory rim shape on the ''wet'' side of the rim (illustrated by
a selection of close-up views shown in the inset). b) Evaluation of
the corresponding rim profiles with regard to the rim width $w$ and
the maximum rim height $H$.} \label{grapha}
\end{figure}

To probe the temporal and spatial growth of the liquid rim,
subsequent \textit{in situ} AFM images during dewetting on a
high-temperature heating plate were recorded in tapping
mode\texttrademark. Cross-sections obtained from a temporal series
of AFM scans (for a time interval $\Delta t=640\,s$) for PS(13.7k)
dewetting from AF\,1600 at 115$\,^{\circ}$C are shown in
Fig.~\ref{grapha}a. The rim clearly grows in height and width as
dewetting proceeds. Nevertheless, the form of the profiles at
different times (and therefore different volumes) is very similar.
All profiles exhibit a substantial trough (see inset of
Fig.~\ref{grapha}a) on the ''wet'' side of the rim, where the height
profile merges into the unperturbed film.

In order to test the self-similarity of the form of the rim, the rim
width $w$ (the distance between three-phase contact line and the point, where the
rim height has dropped to 110\% of the initial film thickness) is plotted as function of the maximum rim height $H$, c.f.\
Fig.~\ref{grapha}b. The linear dependence indicates that
self-similarity can safely be assumed and $w\propto \sqrt{R}$
(where $R$ denotes the radius of the hole) is obtained from the
conservation of volume for growing holes (see Ref.~\cite{Bae10}).

\subsubsection{\textit{Ex situ} Measurement and Analysis of the Rim
Shape at Fixed Hole Radii} \label{exsitu}

By AFM, the rim profile can be recorded at elevated temperature, as
described before, when the PS is above the glass transition
temperature and liquid, or, for simplicity of handling the samples,
at room temperature (\textit{ex situ}). The quenching from the
liquid to the glassy state has in no case lead to differences in the
form of the rim profile. It can be analyzed as described before
\cite{Bae09} (to be more precise, we have used in the following the
third-order Taylor expanded model based on the full Stokes equations
since it applies to a non-restricted range of slip lengths, in
contrast to e.g.\ the strong-slip lubrication model). For reasons of
comparability, all profiles were recorded at a hole radius of
12\,$\mu$m to ensure that approximately the same amount of liquid
is collected in each of the rims.

\begin{figure}[b]
\begin{center}
\includegraphics[width=0.472\textwidth]{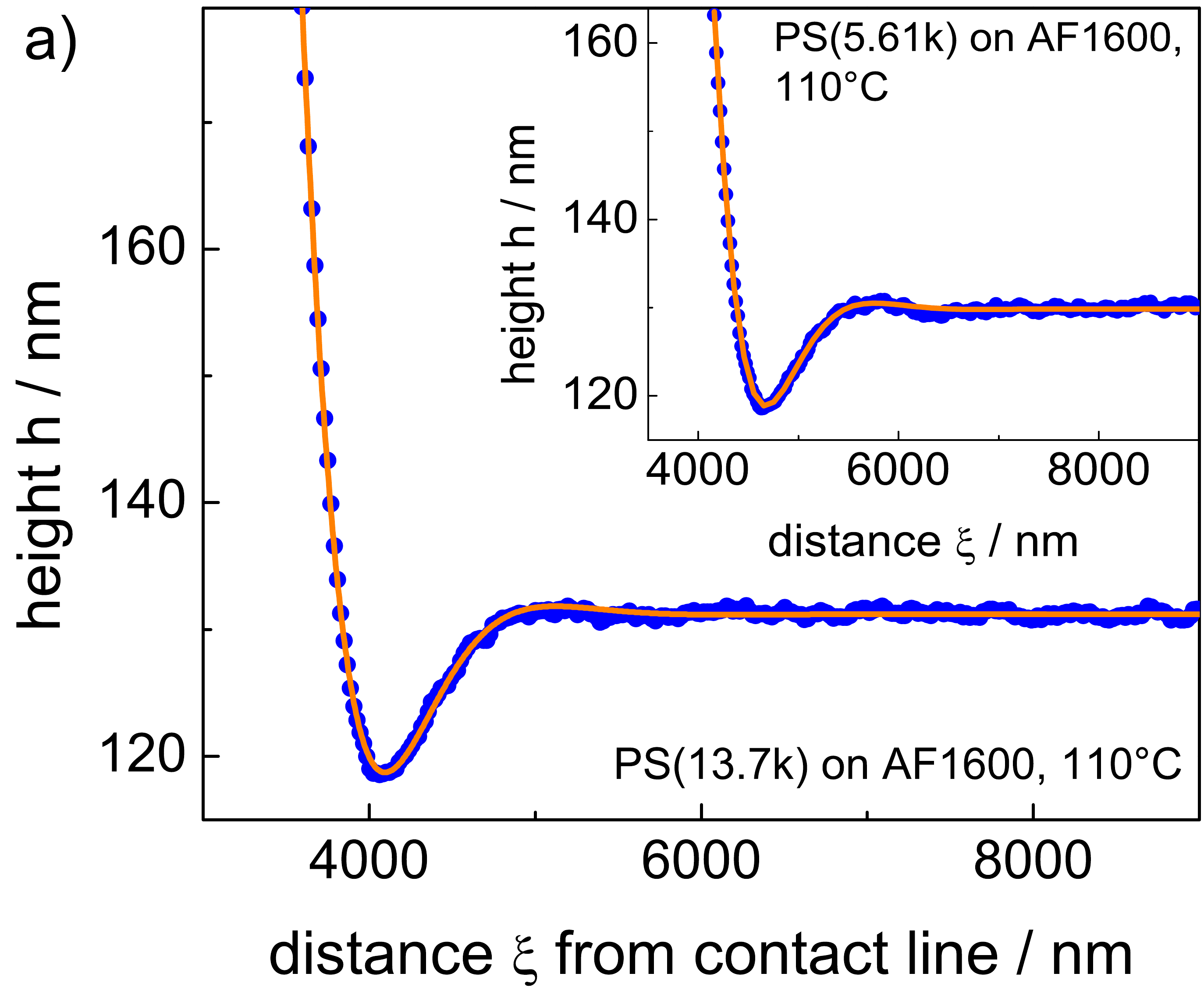}
\includegraphics[width=0.48\textwidth]{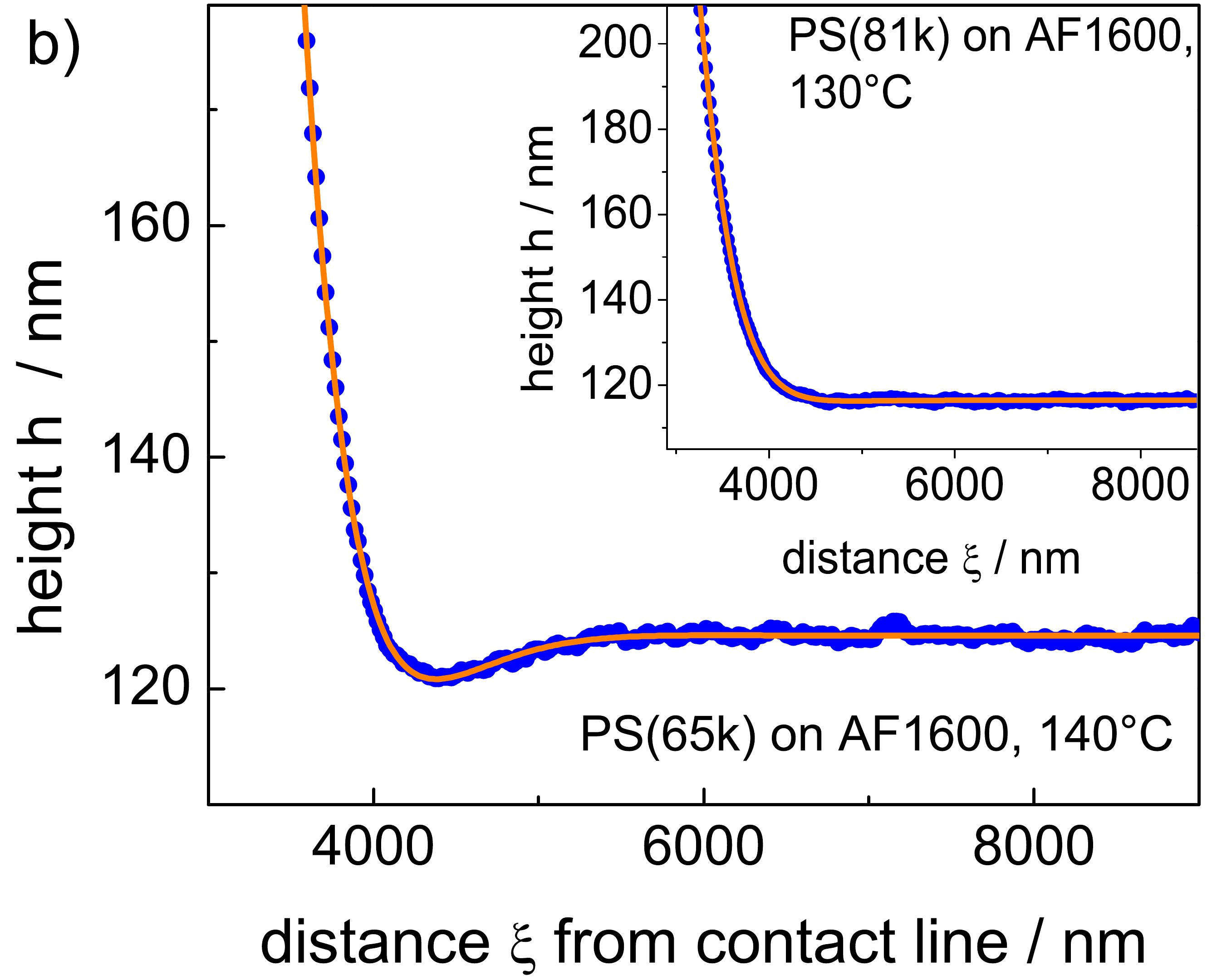}
\includegraphics[width=0.48\textwidth]{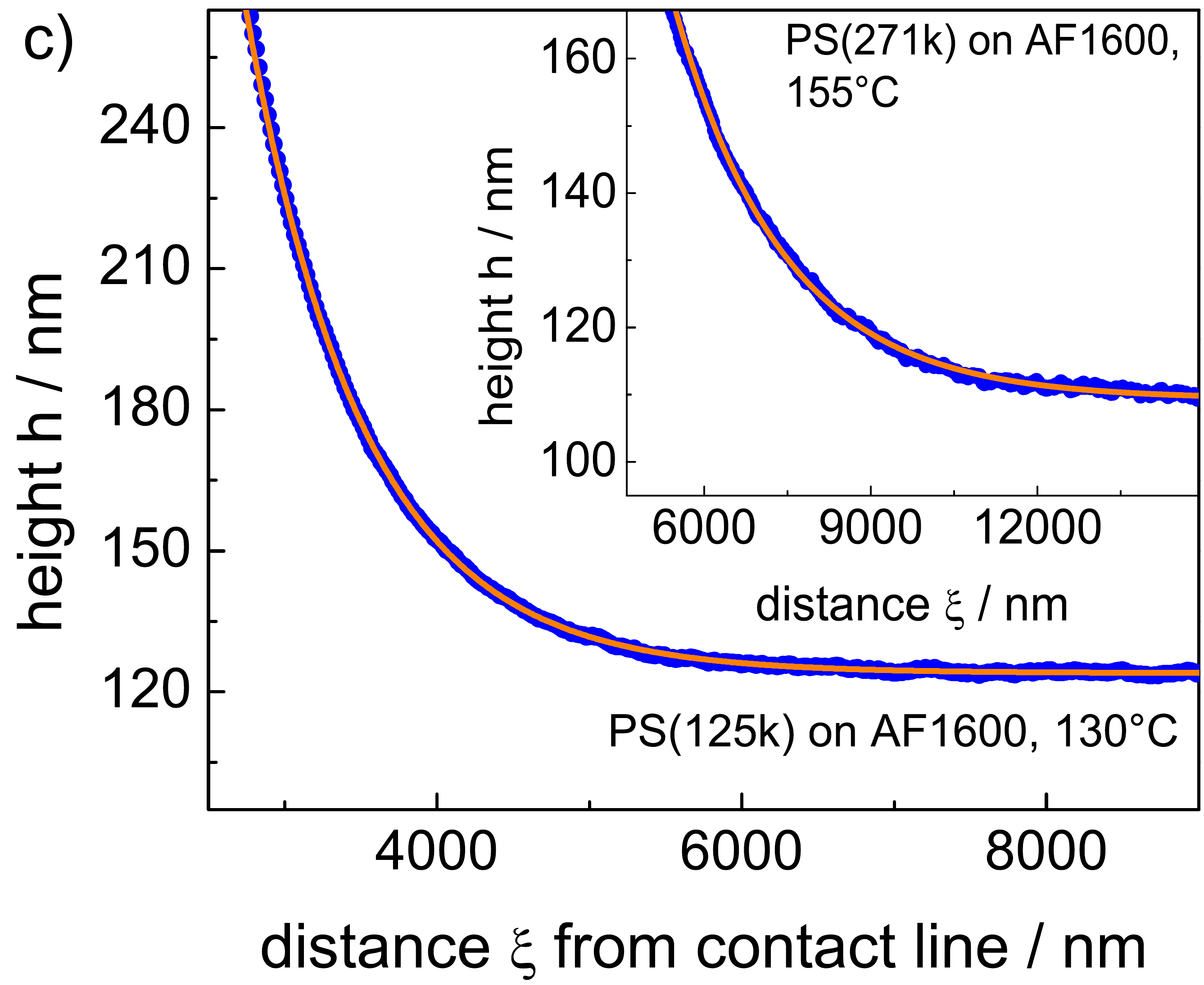}
\end{center}
\caption{Experimentally obtained AFM images of the ''wet side'' of
rim profiles for thin PS films of different molecular weight on
AF\,1600. a) PS(13.7k) and PS(5.61k) (see inset) at 110$\,^{\circ}$C
represent typical rim shapes below the critical molecular weight
$M_\mathrm{c}$ for chain entanglements. b) PS(65k) at
140$\,^{\circ}$C and PS(81k) at 130$\,^{\circ}$C (see inset) are
clearly above $M_\mathrm{c}$. c) PS(125k) at 130$\,^{\circ}$C and
PS(271k) at 155$\,^{\circ}$C (see inset). The solid red lines
indicate the fits to the experimental data according to a damped
oscillation or a single exponential decay.} \label{graphu}
\end{figure}

\begin{figure}[b]
\begin{center}
\includegraphics[width=0.508\textwidth]{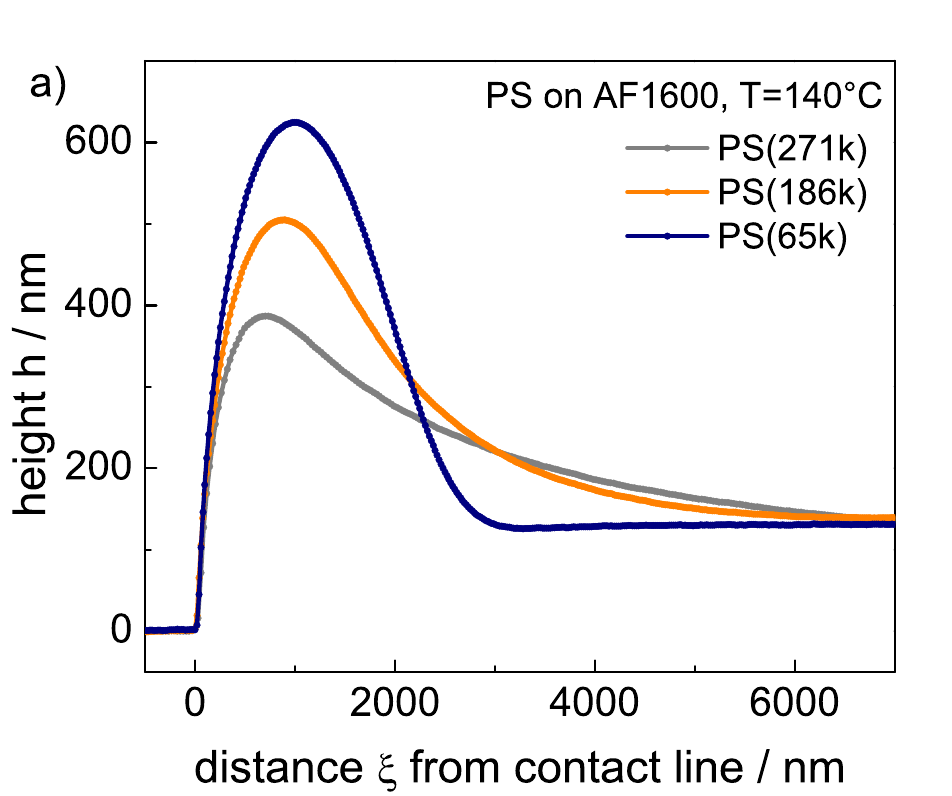}
\includegraphics[width=0.483\textwidth]{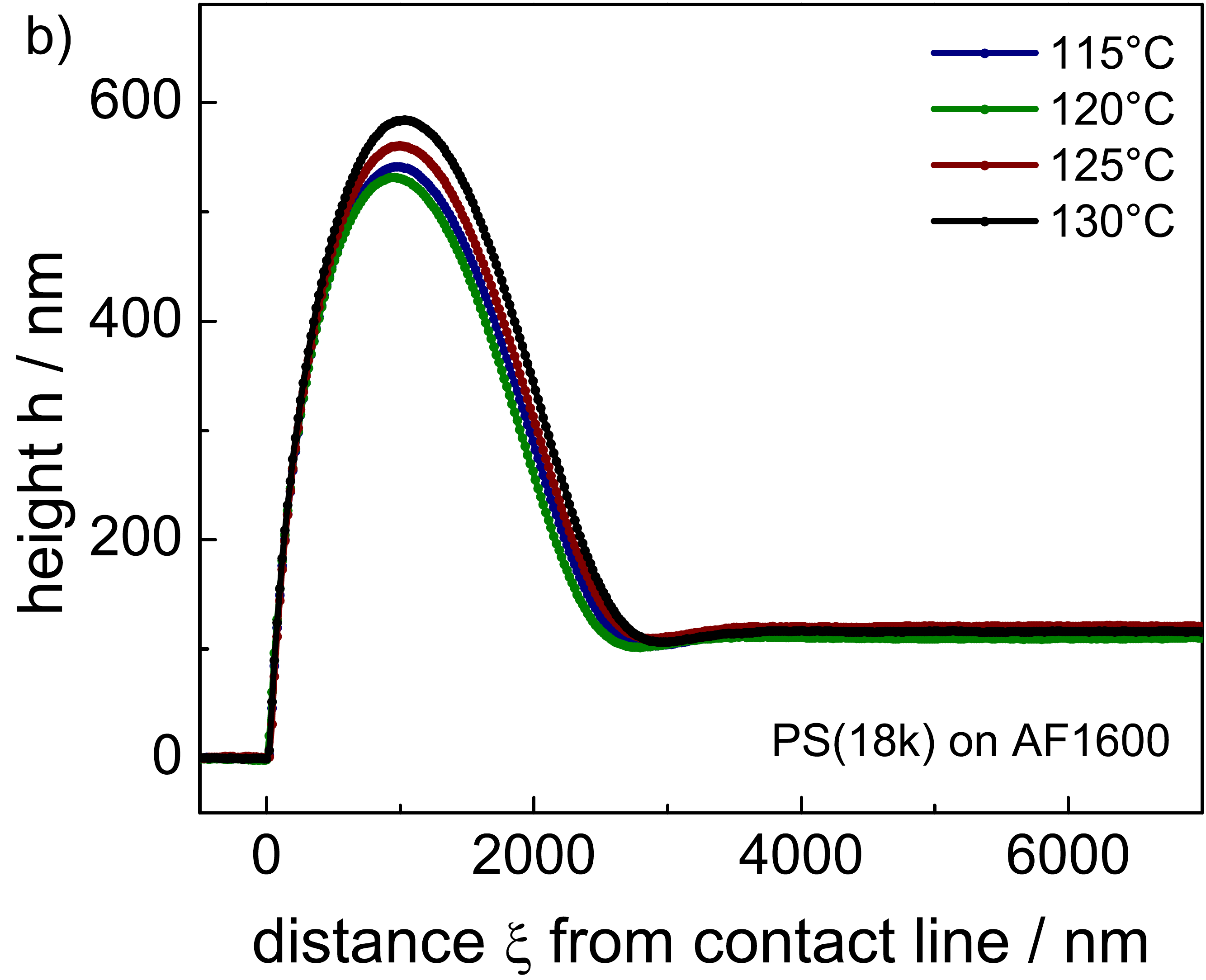}
\end{center}
\caption{a) Examples of rim profiles on AF\,1600 for different
molecular weights $M_\mathrm{w}$. Increasing $M_\mathrm{w}$ results
in a transition from oscillatory to monotonically decaying rims and
to successively less steep decays into the unperturbed film. b) Rim
profiles for PS(18k) as a representative for the typical situation
below the critical chain length for the formation of entanglements.
Dewetting temperature and, thus, viscosity of the polymer film have
no significant impact on the rim profiles (see also
Fig.~\ref{graphp}) \cite{significant}.} \label{grapho}
\end{figure}

\begin{figure}[b]
\begin{center}
\includegraphics[width=0.48\textwidth]{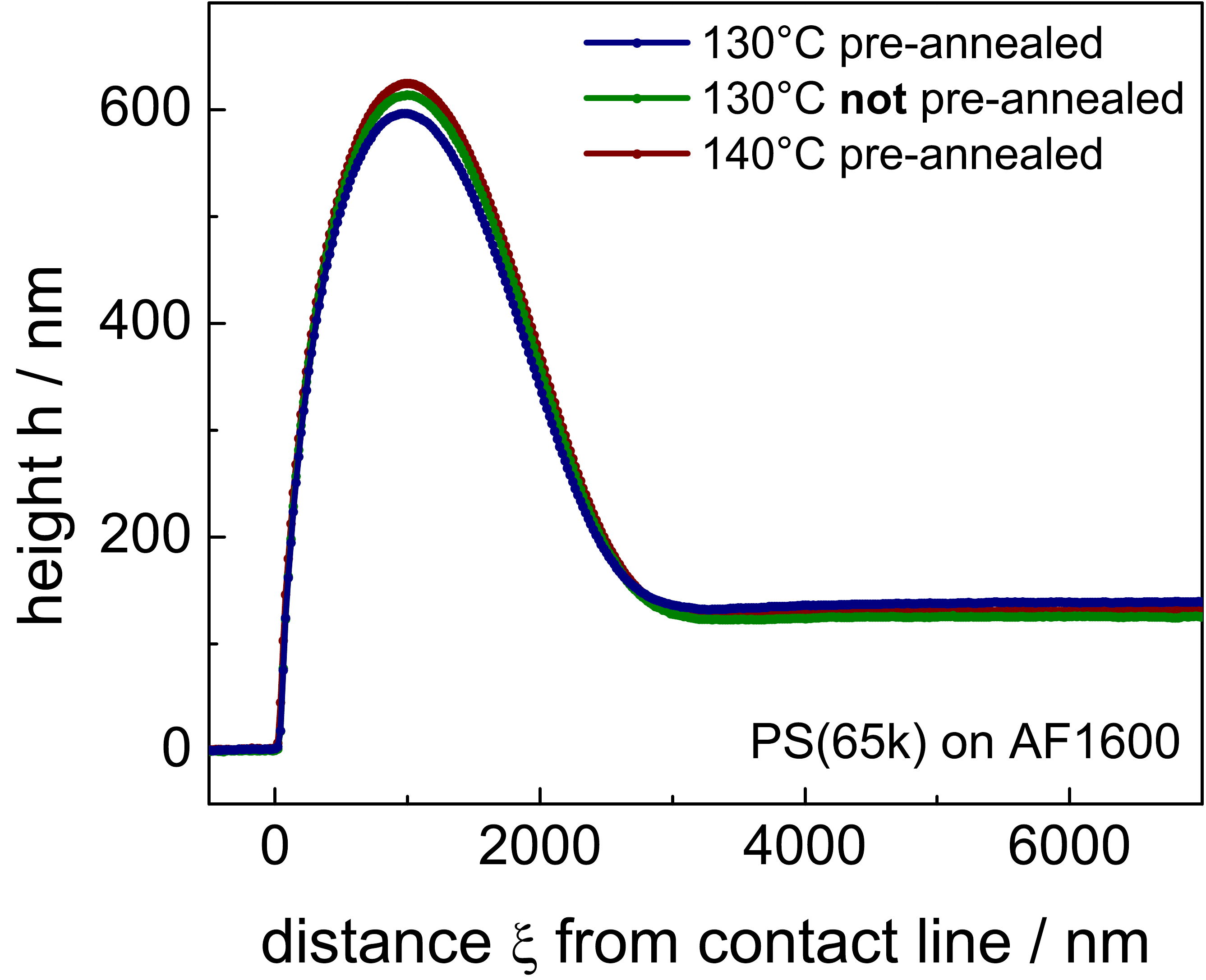}
\includegraphics[width=0.48\textwidth]{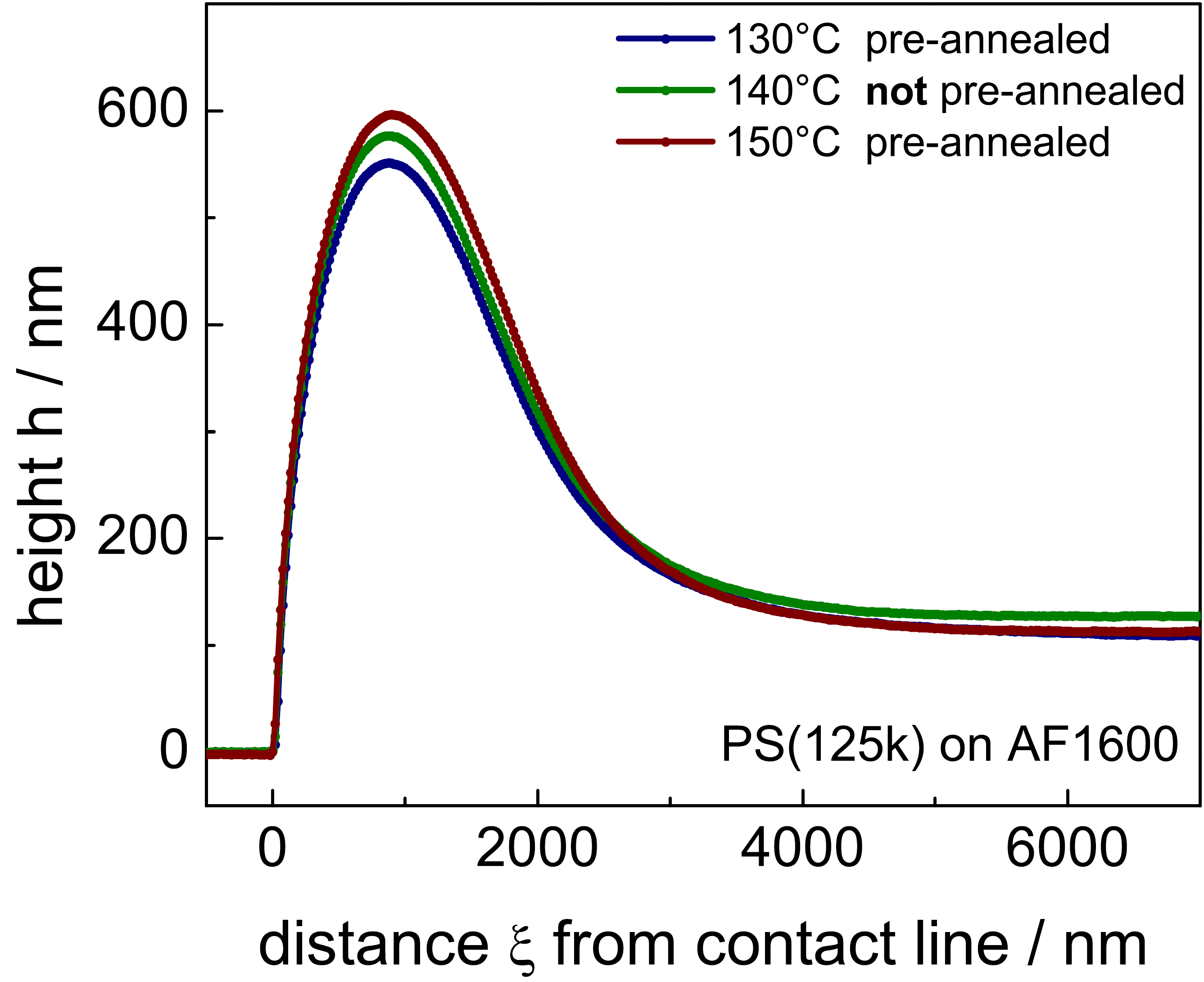}
\includegraphics[width=0.48\textwidth]{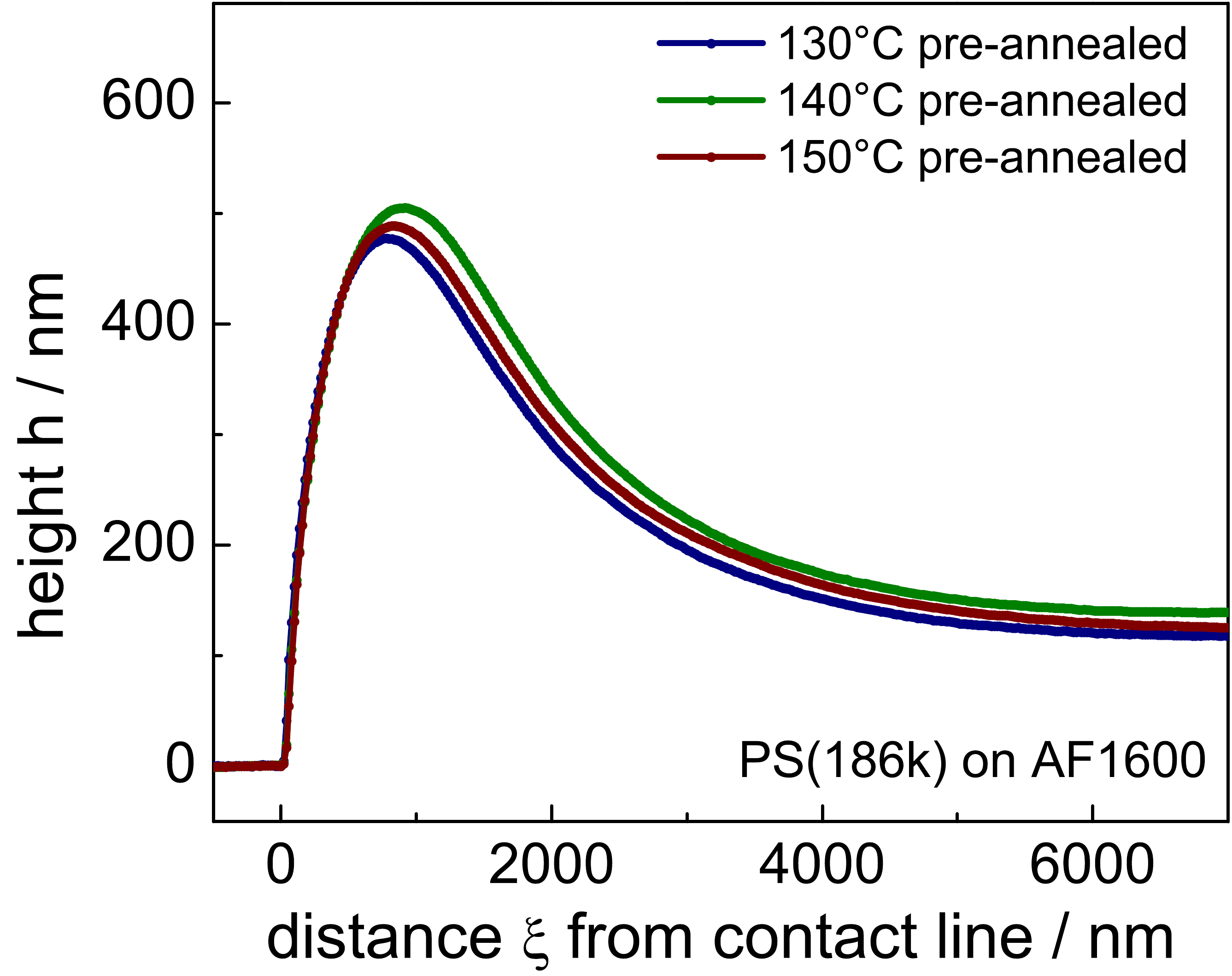}
\includegraphics[width=0.48\textwidth]{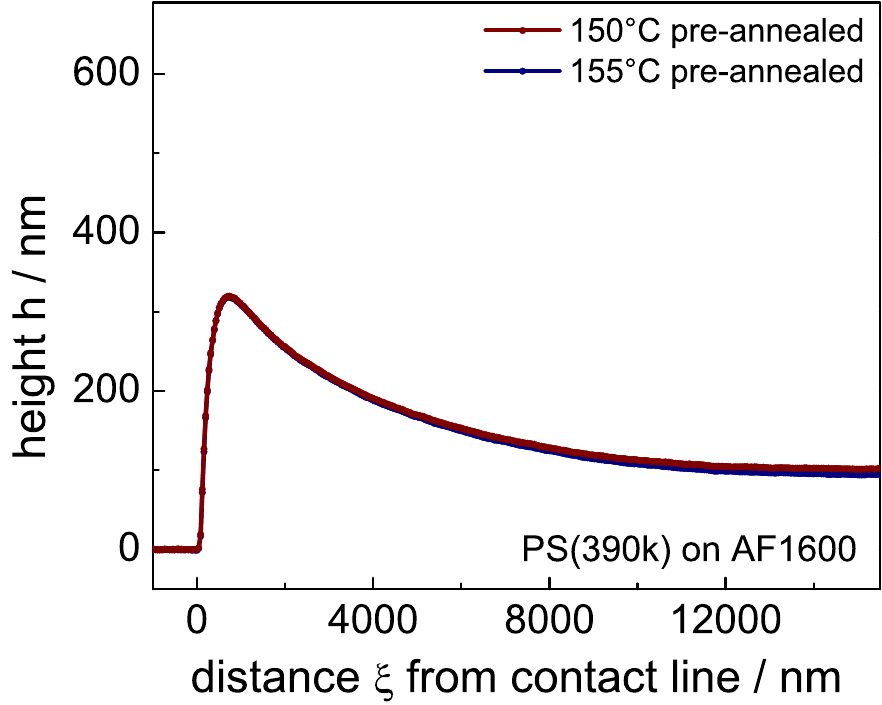}
\end{center}
\caption{Testing the role of pre-annealing (films with or without pre-annealing for 3\,h at 140$^{\circ}$C on mica prior to the transfer to the AF\,1600 substrate) and dewetting temperature for the shape of the rim profile on AF\,1600 for PS(65k), PS(125k), PS(186k) and PS(390k) \cite{significant}.} \label{graphp}
\end{figure}

Rim profiles exhibit two distinct morphological shapes: a) an
oscillatory and b) a monotonic decay.  Exemplarily, Fig.~\ref{graphu} shows typical portions of oscillatory
and monotonically decaying rim shapes and the corresponding fit
functions (exponentially damped sinusoidal oscillation and
exponential decay, respectively) to the experimental data for
various molecular weights. The oscillatory decay, which
is characterized by a significant trough where the rim merges into
the unperturbed film (''wet side''), is illustrated in
Fig.~\ref{graphu}a. In some cases, the experimental data for this
part of the rim profile even exhibits a slight overshoot as compared
to the initial film thickness (see inset of Fig.~\ref{graphu}a).
With increasing $M_\mathrm{w}$, the trough becomes progressively
smaller (see Fig.~\ref{graphu}b). Finally, it vanishes and the rim
exhibits a monotonically decaying profile (see Fig.~\ref{graphu}c). If two exponential decay lengths can be
separated from each other (which is always the case for a damped
oscillation where the decay lengths are a pair of two complex
conjugate), the quantification of the slip length $b$ and of the
capillary number $Ca$ (and also of the viscosity $\eta$ by measuring
the instantaneous dewetting velocity $\dot s$ prior to quenching the
sample to room temperature) is possible. The determination of $b$ in
the case of a single exponential decay, however, necessitates the
knowledge of $Ca$ and, therefore, also of the viscosity $\eta$. The
theoretical models and the application of the fit functions to the
rim profile data are discussed in detail in Ref.~\cite{Bae09}. As
illustrated in Fig.~\ref{graphu}a-c and especially Fig.~\ref{grapho}a, profiles show progressively flatter declines for increasingly larger molecular weights $M_\mathrm{w}$: The rim significantly widens at the expense of a decrease in maximum rim height, an observation also reported in Ref.~\cite{See01}. For a given $M_\mathrm{w}$, however, no systematic dependency on the dewetting temperature and, thus, on the viscosity of the polymer melt can be recorded (see Fig.~\ref{grapho}b and Fig.~\ref{graphp}) \cite{significant}.

To check the influence of possible internal stress in the polymer
film due to the preparation process \cite{Pod01,Rei05,Dam07}, the
films prepared on mica were annealed prior to the transfer to the
AF\,1600 substrate (c.f.\ section~\ref{Preparation}). A comparison of profiles with and without
pre-annealing was exemplarily tested for two molecular weights,
PS(65k) and PS(125k), and is shown in Fig.~\ref{graphp}. For our
systems, we could not detect a systematic influence of pre-annealing
on the form of the rim profile for a fixed molecular weight (see
also section~\ref{sliplength}) \cite{significant}.

\subsubsection{Evaluation of the Slip Length} \label{sliplength}

\begin{figure}[b]
\begin{center}
\includegraphics[width=0.5\textwidth]{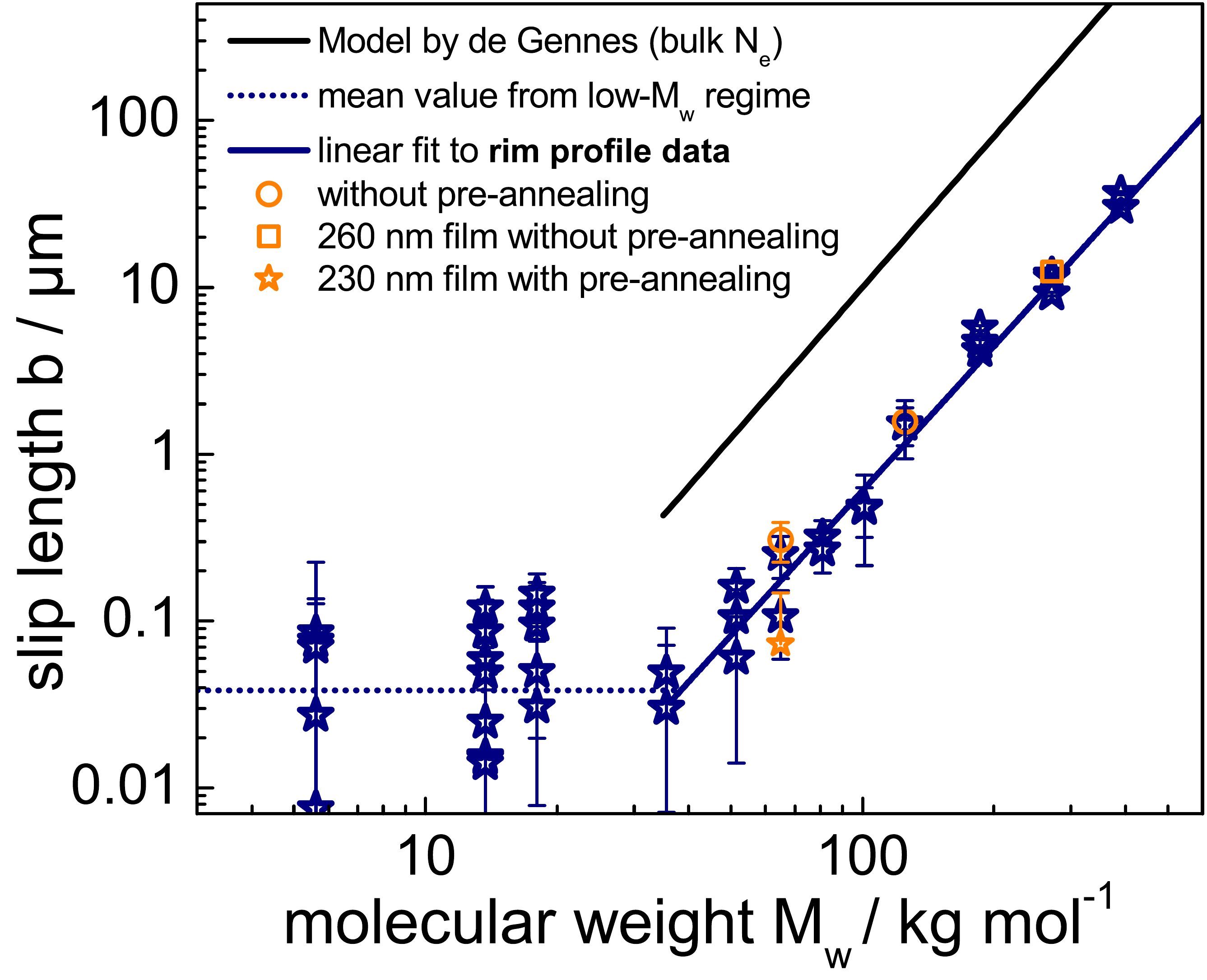}
\end{center}
\caption{Double-logarithmic plot of the slip length $b$ as a
function of molecular weight $M_\mathrm{w}$. $b$ is obtained from
the analysis of rim profiles for single experiments (symbols) with
PS dewetting from AF\,1600 substrates. The solid black line
represents the model by de Gennes \cite{deG79} using bulk
$N_\mathrm{e}$=163, whereas the solid red line is a fit to the
experimental data. Additionally, film thickness and pre-annealing
conditions were varied.} \label{graphk}
\end{figure}

Fig.~\ref{graphk} depicts the results of the slip length $b$ in view
of the molecular weight $M_\mathrm{w}$ in a double-logarithmic
representation: Below the critical molecular weight for chain
entanglements $M_\mathrm{c}$, the slip length reaches values up to
150\,nm at maximum. For $M_\mathrm{w}>M_\mathrm{c}$, the slip
length strongly increases with the molecular weight or the length
of the polymer chain, respectively. To be more precise, a scaling
exponent of 2.9(2) between the slip length and the molecular weight,
i.e.\ $b\propto M_\mathrm{w}^{\beta}$ with $\beta=2.9(2)$, is found.
This result is in excellent agreement with the outcome of the hole
growth dynamics (see section~\ref{holeslip}). The values for $b$
are systematically lowered compared to the expectation by de Gennes
(i.e.\ $b=aN^3/N_\mathrm{e}^2$, see Ref.~\cite{deG79}) based on
literature values for the molecular size $a=3{\AA}$~\cite{Red94} and
the bulk entanglement length $N_\mathrm{e}=163$~\cite{Rub03}. The
pre-factor $a/N_\mathrm{e}^2$ is obtained from the linear fit to the
experimental data in the double-logarithmic representation: For
fixed $a=3{\AA}$, the rim profile analysis leads to
$N_\mathrm{e}=517$, which is a factor of 3.2 larger compared to the
bulk entanglement length of 163 monomer units \cite{a}.

\subsubsection{Checks of Consistency and Validity}

The general consistency of this method to extract the slip length
and the capillary number from liquid rim profiles was checked for
PS(13.7k) dewetting from OTS and DTS substrates in our previous
publications \cite{Fet07,Bae09}: The viscosity (which was determined
from the capillary number and the dewetting velocity) turned out be
in excellent agreement with independent (bulk) viscosimeter data.
\textit{In situ} measurements did prove that the slip length was
independent of the radius of the holes (the range between 2.5 and
17\,$\mu$m was tested). Dewetting experiments for various film
thicknesses (between 50 and 230\,nm) also lead to consistent
results. The method, consequently, yields reliable results for the
slip length and the viscosity of dewetting thin polymer films on the silanized Si wafers. In the following, the identical consistency and validity checks are performed for the dewetting experiments on the AF\,1600 substrate.

\paragraph{Determination of the PS Film Viscosity}

\begin{figure}[b]
\begin{center}
\includegraphics[width=0.5\textwidth]{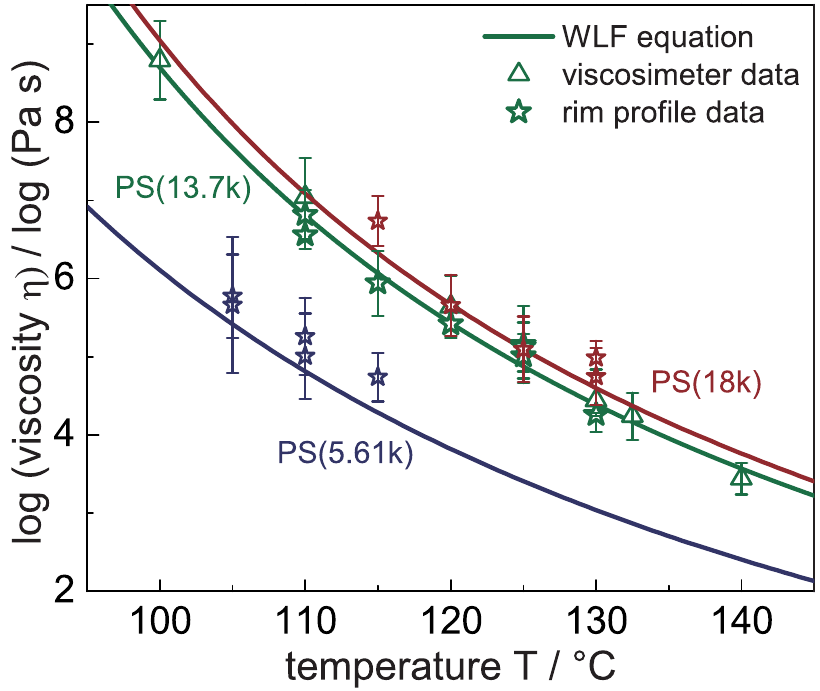}
\end{center}
\caption{Semi-logarithmic plot of viscosity $\eta$ versus
temperature $T$ for PS melts of various molecular weights
$M_\mathrm{w}<M_\mathrm{c}$. The viscosity data is obtained from rim
profile analysis of PS(5.61k), PS(13.7k) and PS(18k) films on
AF\,1600. The solid curves are based on the extrapolation of a fit
to experimental viscosimeter data for PS(12.5k) according to the WLF
equation.} \label{graphvisco}
\end{figure}

For AF\,1600 substrates, the capillary number $Ca$ is extracted from
the analysis of the experimentally recorded rim profiles up to
PS(81k). The knowledge of the surface tension of the polymeric
liquid ($\gamma_\mathrm{lv}$(PS)\,=\,30.8\,mN/m, \cite{Bra99}) and
the experimental determination of the actual dewetting velocity
$\dot s$ (from optical hole growth measurements) enable the
calculation of the viscosity $\eta$ of the polymer melt via
$Ca=\eta\dot s/\gamma_\mathrm{lv}$ for a given dewetting
temperature. Dewetting experiments were performed for various
molecular weights of PS melts below and above the critical molecular
weight $M_\mathrm{c}$ for entanglements. The results are given in
Fig.~\ref{graphvisco} and Tab.~\ref{tabrimviscosity}.
Within the experimental error, good agreement with independent
viscosimeter measurements and their extrapolation according to the
WLF equation \cite{seemann,WLF,Wil55,Rub03} is achieved. From monotonically
decaying rims only one decay length can be obtained. Therefore, the capillary number and, thus, the viscosity cannot be extracted for PS(101k) and larger
$M_\mathrm{w}$ on the AF\,1600 substrate.

\begin{table}[t]
\begin{center}
\caption{\label{tabrimviscosity}PS on AF\,1600: Quantitative
comparison of viscosities $\eta_\mathrm{rim}$ determined from
capillary numbers which were obtained by the analysis of rim
profiles to viscosities $\eta_\mathrm{WLF}$ calculated from the WLF
equation.}
\begin{tabular}{c|c|c|c}
$M_\mathrm{w}$ &  $T$ ($^{\circ}$C) & $\eta_\mathrm{WLF}$ (Pa s) & $\eta_\mathrm{rim}$ (Pa s) \\
\hline
35.6k & 120 & $2.31 \times 10^6$ & $3.5(13) \times 10^6$  \\
35.6k & 130 & $1.61 \times 10^5$ & $2.8(11) \times 10^5$  \\
51.5k & 125 & $1.95 \times 10^6$ & $5.5(26) \times 10^6$  \\
51.5k & 130 & $5.62 \times 10^5$ & $1.8(5) \times 10^6$  \\
51.5k & 140 & $7.11 \times 10^4$ & $1.6(6) \times 10^5$  \\
65k   & 130 & $1.30 \times 10^6$ & $1.7(4) \times 10^6$  \\
65k   & 140 & $1.60 \times 10^5$ & $2.4(6) \times 10^5$  \\
81k   & 130 & $2.63 \times 10^6$ & $1.0(2) \times 10^6$  \\
81k   & 140 & $3.32 \times 10^5$ & $1.7(5) \times 10^5$  \\
\end{tabular}
\end{center}
\end{table}

\paragraph{Preparation Conditions and Non-Newtonian Effects}

In the context of the results of this dewetting study concerning the
flow dynamics of thin polymer films, a discussion of the influence
of potential viscoelastic effects is necessary. A source of
viscoelastic effects are residual stresses caused by the fast
solvent evaporation during the spin-coating process. To safely
exclude this issue, dewetting experiments were typically performed
after a pre-annealing step on the mica substrate (c.f.\ section~\ref{Preparation}). However, to test the influence of this
procedure, additional experiments were carried out without annealing prior to dewetting. Rim profiles of these as-cast films were analyzed
and compared to pre-annealed samples: No systematic influence of the
pre-annealing step is found for the typical molecular weights of
PS(65k) and PS(125k) used in this study (see Fig.~\ref{graphp}). The
slip lengths determined from these experiments are still roughly one
order of magnitude lower than expected from literature values for
$N_\mathrm{e}$ (see Fig.~\ref{graphk}). Consequently, for ''mature'' holes exhibiting fully developed rims, the flow dynamics and morphologies are not significantly affected by polymer relaxation processes that could potentially originate from thin film preparation and residual stresses.

\begin{table}[b]
\begin{center}
\caption{\label{tabWi}Calculated relaxation times $\tau=\eta/G$ and
Weissenberg numbers \textit{Wi} for diverse temperatures $T$ and
molecular weights $M_\mathrm{w}$ of the PS melt dewetting from the
AF\,1600 substrate. The viscosity $\eta$ was obtained from the WLF
equation; $G\approx$\,0.2\,MPa was taken from literature
\cite{Rub03}. The maximum shear rates $\dot\gamma_\mathrm{max}$ and
the maximum values for corresponding Weissenberg numbers \textit{Wi}
were calculated as described in Ref.~\cite{Bae092}.}
\begin{tabular}{c|c|c|c|c||c|c|c|c|c}
$M_\mathrm{w}$ &  $T$ ($^{\circ}$C) & $\tau$ (s) & $\dot\gamma_\mathrm{max}$ (1/s) & $Wi$ & $M_\mathrm{w}$ &  $T$ ($^{\circ}$C) & $\tau$ (s) & $\dot\gamma_\mathrm{max}$ (1/s) & $Wi$ \\
\hline
35.6k & 120 & 11.55 & 0.0045 & 0.052 & 125k & 130 & 57.5  & 0.0003 & 0.019 \\
35.6k & 130 & 0.81  & 0.0546 & 0.044 & 125k & 150 & 1.39   & 0.0167 & 0.023 \\
51.5k & 125 & 9.75  & 0.0023 & 0.022 & 186k & 130 & 222   & 0.0002 & 0.052 \\
51.5k & 130 & 2.81  & 0.0065 & 0.018 & 186k & 140 & 28.05  & 0.0012 & 0.034 \\
51.5k & 140 & 0.36 & 0.0863 & 0.031 & 186k & 150 & 5.4   & 0.0072 & 0.039 \\
65k & 130   & 6.5  & 0.0074 & 0.048 & 271k & 140 & 100.5 & 0.0006 & 0.057 \\
65k & 140 & 0.8    & 0.0368 & 0.029 & 271k & 150 & 19.35  & 0.0040 & 0.077 \\
81k & 130 & 13.15   & 0.0092 & 0.121 & 271k & 155 & 9.55   & 0.0041 & 0.039 \\
81k & 140 & 1.66    & 0.0647 & 0.108 & 390k & 150 & 66.5  & 0.0004 & 0.029 \\
101k & 130 & 27.8  & 0.0037 & 0.103 & 390k & 155 & 32.9  & 0.0018 & 0.058 \\
101k & 140 & 3.52   & 0.0197 & 0.069 &  &  &  &  \\
\end{tabular}
\end{center}
\end{table}

Further the question arises whether the flow itself is Newtonian, i.e.\, whether the viscosity is independent of the shear rate within the range of shear rates present. Weissenberg numbers $Wi=\tau\dot{\gamma}$ smaller than 1 assure that viscoelastic effects of the polymer melt can safely be neglected \cite{weissenberg}. The present experimental configuration, given by the dewetting temperatures and the molecular weights of the polymer, results in characteristic relaxation times, shear rates, and Weissenberg numbers for which estimates  can be calculated (see Tab.~\ref{tabWi}): Relaxation times $\tau$ are determined via the viscosity $\eta$, obtained from the WLF equation (c.f.\
Refs.~\cite{seemann,WLF,Wil55,Rub03}), and the shear modulus $G$ of
the polymer melt. Shear rates $\dot\gamma$ were estimated as described in Ref.~\cite{Bae092} resulting in an upper limit for the corresponding Weissenberg numbers \textit{Wi}=$\tau\dot\gamma$. As indicated in
Tab.~\ref{tabWi}, Weissenberg numbers calculated
from experiments on the AF\,1600 substrate do not exceed a value of
0.1. Consequently, the occurrence of non-Newtonian
effects is not expected for the
examined molecular weights and the low shear rates (see
Tab.~\ref{tabWi}) in such dewetting experiments. The analysis of the
dewetting experiments presented in this study can therefore safely be based on
the assumption of Newtonian flow.

\paragraph{Impact of the Film Thickness}

Finally, the question arises whether slippage of a thin liquid film, representing fluid motion in a confined geometry,
depends on the level of confinement. To test the effect of the
film thickness on the slip length, an additional dewetting experiment (without
pre-annealing) was performed using PS(271k) at 140$\,^{\circ}$C with
a film thickness of 260\,nm. The results are presented within Fig.~\ref{graphk} as the red data points: The analysis of the rim profile of this hole
gave a slip length of 12.4\,$\mu$m. This is, within the experimental
error, in very good agreement with the corresponding slip length of
11.8\,$\mu$m for a 115\,nm thick PS(271k) film. A further hole
growth experiment in a 230\,nm PS(65k) film (after a typical pre-annealing of the film on mica) 
was recorded at 130$\,^{\circ}$C. The analysis of the corresponding
rim profile exhibited a slip length of 73\,nm. Both experiments
using thicker films do not show a significant increase in the slip length.


\section{Discussion} \label{discussion}

In literature, slip lengths determined from molecular dynamics (MD)
simulations in non-wetting conditions are reported to be in the
range of 15 molecular diameters \cite{Bar991}. Approximating the
molecular diameter by twice the radius of gyration $R_\mathrm{g}$ of
a polymer coil (for PS(13.7k): $R_\mathrm{g}=3.2$\,nm) results in
values for the slip length of up to 100\,nm. From this point of
view, the experimental results for the slip length on AF\,1600 below
$M_\mathrm{c}$ are in good accordance with the theoretical
expectation.

For molecular weights above $M_\mathrm{c}$, $b \propto
M_\mathrm{w}^3$ is found in accordance with de Gennes' prediction.
However, using the bulk value for the entanglement length,
$N_\mathrm{e}=163$, slip lengths are expected to be roughly one
order of magnitude larger than what is found in our experiments. The
deviations can be explained by a reduced entanglement density
compared to bulk values. This is quantified by an increased
entanglement length of $N_\mathrm{e}$=517. 

According to the fact that slip lengths are reproduced for markedly thicker films, we conclude that the observed reduction of the entanglement density is not caused by the confinement of the film but is localized at the solid/liquid interface. The lack of a significant temperature dependence highlights the fact that slippage and flow dynamics are directly related to inter-chain entanglements rather than viscosity: The difference in chain length between e.g.\ PS(51.5k) at $T$=125$\,^{\circ}$C and PS(271k) at $T$=155$\,^{\circ}$C is responsible for a increase in slip of two orders of magnitude, while the viscosity ($2\times10^6$\,Pa~s) is identical for the two examples \cite{Bae092}.

\subsection{Reduced Interfacial Entanglement Density and its
Implications}

\begin{figure}[b]
\begin{center}
\includegraphics[width=0.4\textwidth]{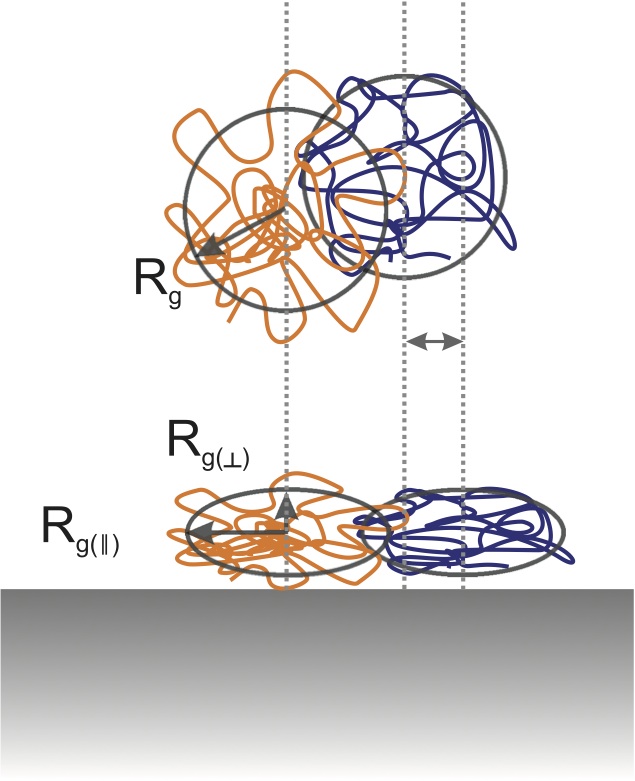}
\end{center}
\caption{Polymer conformation in a bulk melt (upper figure) and at
the substrate/melt interface (lower figure): Two neighboring coils
interpenetrate each other and may form entanglements. At the
interface, the radius of gyration parallel to the substrate
$R_\mathrm{g(||)}$ is comparable to bulk $R_\mathrm{g}$, while it
is reduced in perpendicular direction ($R_{\mathrm{g}(\bot)}$).
Consequently, the coil-coil interpenetration decreases resulting in
a loss of entanglement at the substrate/melt interface. Inspired by Ref.~\cite{Sol08}.}
\label{smarties}
\end{figure}

The analysis of the slip length as a function of the molecular
weight of the polymer has shown that close to the interface,
deviations from bulk properties are to be expected. Polymer chain
entanglements and hence the entanglement length of the polymer are
related to the packing of the polymer chains in the bulk. Close to
an interface, however, the chains pack differently. The reduced
entanglement density in the vicinity of the solid/liquid interface
results from a reduction in the volume pervaded by a polymer chain
at an interface as described by Brown and Russell \cite{Bro96}: The
polymer coil cannot cross the boundary at the interface that rather
acts as a reflecting plane, as previously proposed by Silberberg
\cite{Sil82,Sil88} and is known as Silberberg's principle. 
Chain segments fold back on the same chain and,
thus, the degree of coil-coil interpenetration decreases (c.f.\
Fig.~\ref{smarties}). The reflection process causes a decrease of
the radius of gyration normal to the plane, but does not affect the
radius of gyration parallel to the plane. The polymer coil can be
imagined having an ellipsoidal rather than a spherical shape
\cite{Kra00}. Close to the interface or the wall, the volume
pervaded by a given chain length is smaller than in the bulk. Brown
and Russell \cite{Bro96} suggest that the vertical distance affected
by this phenomenon can be approximated at most by the radius of
gyration $R_\mathrm{g}$ of the polymer. Hence, the normal bulk entanglement density
is expected to be recovered at a distance of $R_\mathrm{g}$ above the interface. They analytically calculate a reduction of the entanglement density or, analogous to that, an increase of the entanglement length by approximately a factor of 4, which is in accordance with the experimental results presented here. 

\begin{figure}[b]
\begin{center}
\includegraphics[width=0.8\textwidth]{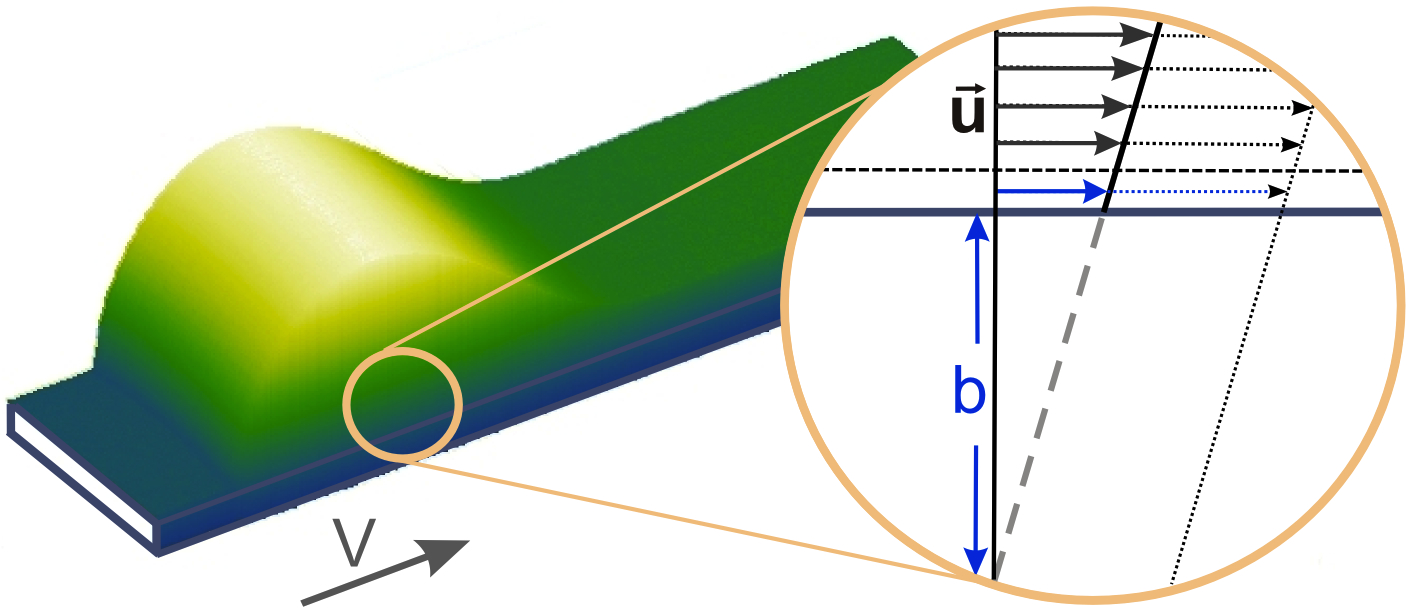}
\end{center}
\caption{AFM height profile of a dewetting rim (dewetting velocity $V$) and (non-scaled) illustration of the flow profile (velocity field $\vec{u}$) for entangled polymer films. The apparent slip velocity (blue arrow), representing the interfacial region at the solid/liquid interface as described in the text, is determined by the local density of inter-chain entanglements. A reduced interfacial entanglement density, compared to the bulk entanglement network, is responsible for a significantly smaller slip velocity and a shift of the entire flow profile towards smaller velocities. The slip length $b$ is obtained as the extrapolation length of the flow profile to zero velocity.} \label{cover}
\end{figure}

The impact of the density of inter-chain entanglements on the flow profile and the slip velocity at the interface of an entangled polymer melt and a solid substrate is difficult to resolve on the molecular level. For a simple liquid, true and apparent slip can be distinguished according to the velocity of the first as compared to subsequent layers of molecules (see e.g.\ Ref.~\cite{Bae10}). In contrast to this, the motion of polymer chains in entangled melts is much more complex, as e.g.\ De Gennes' reptation model shows. To interpret our results, we propose an apparent slip velocity that depends on the entanglement density within this interfacial layer. This picture does not take into account relative motion on the segmental level within this region and, therefore, cannot answer the question whether the monomers or even the chain segments that are in contact with the substrate exhibit a non-zero velocity. The apparent slip velocity, representing the entire interfacial region, shifts the flow profile within the liquid film towards larger velocities. In the strong-slip situation that we are reporting here for entangled polymers, the flow profile is linear and its gradient is unaffected by slippage as it depends on the viscosity of the (bulk-like) residual film. Fig.~\ref{cover} illustrates the difference between apparent slip for a reduced interfacial entanglement density and for a bulk-like entanglement network.

Recent experimental studies probing the effect of confined polymer
chains in thin films have also provided strong evidence of
deviations from polymer bulk properties and a reduced entanglement
density: Si \textit{et al.} report a decrease of the entanglement
density in thin freestanding films compared to the bulk polymer
\cite{Si05}. They conclude that the total entanglement density is
constant, but due to the reduced chain overlap near interfaces, the
proportion of self-entanglements increases at the expense of
inter-chain entanglements. Rowland \textit{et al.} recently studied
the squeeze flow of entangled polymers in confined conditions of
films thinner than the size of the bulk macromolecule \cite{Row08}.
Their experiments reveal a reduced interpenetration of neighboring
chains in favor of self-penetration leading to a weakening of the
polymer network and a decreased resistance to melt flow due to the
presence of less inter-chain entanglements. Molecular dynamics simulations and corresponding primitive path analysis for polymers studying chain conformations and local dynamics corroborate the occurrence of a reduced entanglement
density near a wall \cite{Vla07,Mey07}. Further studies address the film preparation process as a source of non-equilibrium conformations: Barbero and Steiner
studied the mobility at the surface, i.e.\ at the liquid/air
interface of thin polymer films by an electric field technique
\cite{Bar09}. They attributed the reduced effective viscosity in the
case of as-cast films compared to annealed ones to a reduced,
non-equilibrium entanglement density resulting from preparation.

The effect of an altered entanglement network and, in particular, a
reduced entanglement density at an interface also impacts
polymer/polymer (and their interdiffusion) or polymer/air
interfaces, for which the hypothesis of the reflection of polymer
segments is applicable, too. The unique feature of the experimental
results presented here consists in the fact that sliding friction is
sensitive to the proximate properties of the polymer at the
solid/liquid interface, though the residual bulk film is not necessarily
affected by confinement. The above mentioned experimental studies \cite{Si05,Row08}
relate their findings to the confinement given by the film thickness
(and its ratio to the spatial dimensions of the polymer chains
therein). Here, however, the direct and unique impact of the
solid/liquid interface upon chain conformations and rheological
properties of the polymers in the vicinity of a substrate or wall is revealed. The
thickness of the layer affected by the reduced entanglement density
at the solid/liquid interface is at most in the range of the radius
of gyration $R_\mathrm{g}$ \cite{Bro96}. $R_\mathrm{g}$ depends on
the chain length of the polymer and varies for typical PS molecular
weights $M_\mathrm{w}>M_\mathrm{c}$ used in this study from 5.2\,nm
for PS(35.6k) up to 17.1\,nm for PS(390k) \cite{Rg}. Moreover, at
the polymer/air interface, a reduced entanglement density might be
present that - potentially - propagates the same distance away from
the interface into the polymer film. For films thicker than 100\,nm,
the ''bulk region'' of the film is even for the largest
$M_\mathrm{w}$ of our study still more than two times larger than the
superposition of both effects (layers). In some studies, however, $R_\mathrm{g}$ is chosen to be larger than the film thickness, leading to effects of ''overlapping interfacial layers''. Hence, these special cases cannot be captured via the methods presented here. These cases are especially dependent on the preparation procedure and show massive effects upon pre-annealing \cite{Rei05,Dam07}.

The decrease of the entanglement density at the solid/liquid
interface is, as described before, inferred from the determination
of the entanglement length $N_\mathrm{e}=517$. This value
corresponds to a molecular weight of the polymer strand between two
entanglements of $M_\mathrm{e}=53.8\,$kg/mol. Thus, being
significantly larger than the critical molecular weight
$M_\mathrm{c}$ found for bulk situations, this consideration
suggests that the formation of entanglements at the interface starts
at a higher molecular weight as compared to the upper residual
(bulk-like) polymer film.

\subsection{Quantitative Comparison of Slip Lengths: Rim Profile
Analysis vs. Hole Growth Dynamics}

Slip lengths obtained from hole growth dynamics agree qualitatively with those gained from the analysis of rim profiles: Slip is independent of the molecular weight for unentangled films, whereas the slip length increases $\propto M_\mathrm{w}^3$ for molecular weights above $M_\mathrm{c}$. 

While results are also in good quantitative agreement for low molecular weights, however, a significant deviation is found for $M_\mathrm{w}>M_\mathrm{c}$: data extracted from hole growth are typically larger by a factor of about 2\,-\,3 compared to data obtained from the rim shapes (c.f.\ Fig.~\ref{graphi}). This difference is potentially caused by an inevitable assumption included in the energy dissipation model of sliding friction: Regarding the two dimensional representation of the rim, the lateral extent of slippage in the $x$-direction, i.e.\ in the direction of dewetting, is approximated by the rim width $w$. The latter is again connected to the slip length: Increasingly larger slip lengths are correlated to wider rims. The rim width, however, has been defined for practical reasons as the distance between the three-phase contact line and the $x$-coordinate at which the rim height has dropped to 1.1 times the initial film thickness $h_\mathrm{0}$. If, however, the ''true'' lateral extent of slippage is significantly smaller than the rim
width $w$, the frictional energy is dissipated over a narrower extent
and the slip lengths obtained from the model for the hole
growth analysis (assuming a superposition of viscous flow and
slippage, c.f.\ Eq.~(\ref{velocityequation})) would be systematically too large. This hypothesis is supported
by the dewetting experiments on the AF\,1600 substrate below the
critical chain length for entanglements: In case of these solely
weakly slipping PS films, no significant difference between slip
lengths obtained from rim profiles and hole growth dynamics is
found. Consequently, the significant shift between the results of
the two methods is either correlated to the change of the rim
profile, in combination with the definition of the rim width, or,
and this is the second possibility to be taken into consideration, to the change of
the velocity profile according to the presence of slippage. Gaining
information about the extent of slippage is difficult. In
section~\ref{FurtherExperiments}, an experimental system will be
introduced that might help to shed light on this issue: thin
block-copolymer films dewetting from hydrophobic substrates.

Besides the aforementioned methodical issue of the differences
concerning the determination of slip lengths from the hole growth
dynamics and the analysis of rim profiles, a further explanation can
be hypothesized: Rim profiles are sensitive to the local
hydrodynamic boundary condition in the region of the solid/liquid interface where the rim decays into the unperturbed film. The hole growth analysis, however, is sensitive to
the entire extent of the rim width $w$.

To sum up the above discussion, the
systematic differences in slip length originating from the two
different techniques may be linked to the lateral extent of slippage
underneath the liquid rim.



\section{Pushing to the Limits - More Sophisticated Dewetting Experiments}
\label{FurtherExperiments}

In the previous section, we addressed the question to what lateral extent the area underneath the rim is exposed to interfacial flow and in particular interfacial frictional forces: Which part of the liquid rim is actually capable to slide over the substrate's surface? To explore this issue, we modified the experimental system and switched to liquids with an internal structure, block-copolymers. The idea is to prepare a well-known di- or triblock-copolymer microphase-separated thin film and then to induce flow by dewetting from a hydrophobic substrate. Similar to homopolymer films, holes break up and grow in time while forming a liquid rim. At distances from the three-phase contact line where the interfacial microphase structure is exposed to shear forces, a structural orientation of the cylindrical micro-domains is expected to be induced; the phenomena of shear-induced orientation has already been reported by other studies before \cite{Zve98,Kno02}. 

\begin{figure}[b]
\begin{center}
\includegraphics[width=0.5\textwidth]{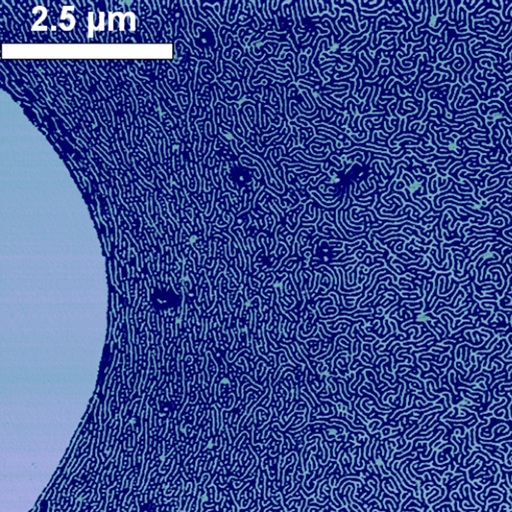}
\end{center}
\caption{Microphase structure (AFM phase image) of a PS/PEP
block-copolymer film at the substrate/polymer interface as recorded
from underneath. The image shows the vicinity of an hole (bright
part on the left side), which resulted from dewetting of the film
from a hydrophobized (OTS) Si wafer at 140$^{\circ}$C.}
\label{PSPEP}
\end{figure}

After quenching the sample to room temperature, the film is detached from the substrate and the rim is imaged from underneath with the AFM. Differences in the orientational structure of the two phases become visible for different lateral displacements from the three-phase contact line \cite{Hae11}: Fig.~\ref{PSPEP} depicts an AFM (tapping mode\texttrademark) phase image of a PS/PEP block-copolymer (Kraton$^{\circledR}$ G1701E, volume percentage of PS: 28\%, Kraton Polymers, Houston, TX, USA) film dewetting from a hydrophobized (OTS-covered) Si wafer. A comparison of the width of the oriented area, as e.g.\ determined from Fourier transforms of consecutive small areas, to the topographical rim width can now be accomplished. Roughly, the width of the oriented area is 1-2\,$\mu$m. The range corresponds to the width of the rim. (Details will be published elsewhere \cite{Hae11}.) Ongoing studies focus on the systematic modification of slip properties by a variation of the type of substrate, which has shown to be strongly correlated to slip of PS homopolymer melts (c.f.\ Fig.~\ref{rversust}), and further parameters such as the dewetting temperature \cite{Hae11}. In case of a strong-slip boundary condition, the structure close to the solid/liquid interface should experience friction rather than significant shear: it is rubbed over the surface. This should affect the microphase structure in a different way as compared to shearing, which is related to a no- or weak-slip boundary condition.

\section{Conclusion and Outlook}

The challenge of explaining facets such as sliding friction and
the fluid dynamics of polymers at the solid/liquid interface on
the molecular level is a main task in the field of micro- and
nanofluidics. The aim of this study was the characterization of
the flow dynamics of thin polymer films in view of the
hydrodynamic boundary condition at the solid/liquid interface.
Dewetting experiments are a powerful tool to study the mechanisms
of friction and the rheological properties of polymers on the
nanoscale.

Hole growth dynamics and rim profile analysis allow for a quantification of the impact of the molecular weight and, in particular, of chain entanglements on the sliding properties of polymers: On smooth, hydrophobic AF\,1600 surfaces, a weak-slip situation is found for PS melts with molecular weights below the critical molecular weight $M_c$ for entanglement formation. This result is independent of the molecular weight in this regime. However, slip is strongly enhanced above $M_c$: The slip length scales with the third power of the molecular weight of the polymer, as predicted by De Gennes. At the interface itself, the density of chain entanglements is found to be reduced by a factor of 3 to 4 compared to the bulk, caused by the different packing of polymer chains at an interface according to Silberberg's principle.

The results are of extraordinary importance in view of a series of
applications such as e.g.\ polymer extruders or, in general,
microfluidic devices. As pointed out by Bocquet and Barrat
\cite{Boc07}, the mean velocity of a pressure drop flow in a
cylindrical channel is increased by a factor $1+8b/h$ as compared
to the classical no-slip boundary condition, where $b$ is the slip
length and $h$ the thickness of the channel. Thus, the
permeability of channels and also of porous media is highly
increased by slippage. An increase in the slip length for longer
polymer chains by two orders of magnitude, as presented in this
study, would also impact the mean velocity in sub-micron channels
by orders of magnitude.

In addition to the influence of polymer chain length on slippage discussed in this paper, we shortly reported on significant dynamical and morphological differences purely resulting from the choice of different hydrophobic substrates (c.f.\ Fig.~\ref{rversust}). In terms of drawing a general molecular picture, not only the degree of entanglement but also structural and dynamic properties on the monomer and segmental level imposed by the substrate have to be discussed. As presented in Tab.~\ref{tabsubstrateproperties}, characteristic substrate properties such as roughness (based on AFM measurements), macroscopically determined wetting properties and surface energies, however, reveal no clear correlation of one of these parameters to slippage. For a discussion of the interfacial structure on the molecular level and its implications on the dynamical properties of the liquid, the reader is referred to our recent study \cite{Gut11}: Neutron and x-ray reflectometry are powerful experimental tools to explore the molecular structure of the solid/liquid interface and add new insights explaining the differences in the sliding properties of PS on different substrates. Currently, also dewetting experiments are extended: The aforementioned methods and theoretical models can also be applied to the spatial and temporal evolution of thin poly(methyl methacrylate) (PMMA) films. The results might shed further light on the link between the monomeric structure of the polymer and slippage at the solid/liquid interface.

Aside from the analysis of the hole growth dynamics and the
morphology of the rim surrounding the hole, recent experiments and
theoretical studies also focus on the liquid rim instability as an
indicator for the slip/no-slip boundary condition
\cite{Mue052,Kin09,Mue11,Mar11}. This instability is comparable to
the Rayleigh-Plateau instability emerging on a surface of a
cylindric jet of water flowing off a tap. The resulting surface
corrugations can be optically observed and the jet finally decays
into discrete droplets. The morphology of the instability of
retracting straight fronts in thin liquid films is governed by the
hydrodynamic boundary condition: While under no-slip conditions
symmetric liquid bulges (co-moving with the front) are found,
slippage causes the formation of asymmetric finger-like
structures. This phenomenon can be optically observed and
represents a further way to access slippage in thin polymer films.
Further characteristics of the instability such as the evolution
of the wavelength of the instability along the receding front (on
lateral scale and perpendicular to the surface plane) are subject
of ongoing investigations \cite{Mar11}.

We hope that our results will motivate
further theoretical studies and simulations, as this phenomenon could
be of high relevance for other experiments such as e.g.\ colloidal
particles moving through a liquid medium.



\section*{Acknowledgments}
The authors acknowledge financial support from the German Science
Foundation (DFG) under grant JA905/3 within the priority program
1164 \textit{''Micro- and Nanofluidics''} and the graduate school
GRK1276. We thank B. Wagner, A. M\"unch, R. Seemann, M. Rauscher and
M. M\"uller for fruitful discussions and R. Seemann for providing
some of the viscosity data.



\begin{thebibliography}{100}
\bibitem{Si_wafer} Si wafers have to be cleaned from hydrocarbon residues which are typical
leftovers from the polishing process. Usually, oxidizing agents are
used like 1:1 mixtures of $H_2SO_4:H_2O_2$. After that it is
essential to rinse the wafers in hot MilliQ\texttrademark \,water to
remove the oxidized substances from the wafer surface. The wafer
surface should be controlled with a reflective light microscope in
dark field illumination, where particles of some nm in size can
still be detected. On a carefully cleaned Si wafer, a water droplet
spreads, the contact angle is zero.
\bibitem{mature} Care has to be taken that the holes have already a certain size,
since the precondition for the validity of the mentioned growth
laws is the self-similarity of the growing rim.
\bibitem{seemann} Viscosimeter data was provided by R. Seemann for PS(12.5k) and PS(65k) for different temperatures.
\bibitem{WLF} The reconstruction of the viscosity $\eta$ as function of temperature
$T$ according to the WLF equation requires the knowledge of the
glass transition temperature
$T_\mathrm{g}=T_\mathrm{g}(\infty)-C/M_\mathrm{w}$ for the
corresponding $M_\mathrm{w}$, the Vogel temperature
$T_\mathrm{Vogel}$ and the ratio of the parameters $B/f_\mathrm{g}$: For PS melts, $T_\mathrm{Vogel}$ (in Kelvin
units) is given by $T_\mathrm{Vogel}=T_\mathrm{g}-48\,$K.
$B/f_\mathrm{g}=30.3$ is found according to literature \cite{Rub03}.
The glass transition temperature $T_\mathrm{g}$ of supported PS films was determined from ellipsometric measurements of the change of the thermal expansion coefficient: $C=70\,$K\,kg\,mol$^{-1}$
and $T_\mathrm{g}(\infty)=372\,$K is found.
\bibitem{AFM-damage} Liquids can be imaged by AFM using non-contact mode. Hereby, care has to be taken to adjust the scan parameters appropriately. By scanning the same area at different scan sizes, possible damages can be inferred. It is also possible to quench a polymer melt to the glassy state. In non-contact mode, our experiments did not reveal differences between rim profiles of glassy and of liquid polymer.
\bibitem{significant} Within one $M_\mathrm{w}$, the rim profiles deviate slightly in maximum rim height, yet neither significantly with the dewetting temperature nor with the pre-annealing procedure. The deviation in height is rather likely caused by small variations of the
prepared film thickness and differences in hole radius which lead to
small variations of the volume of the rim.
\bibitem{a} To the best of our knowledge, a smaller value for the molecular size than
$a=3{\AA}$, \cite{Red94}, is not reasonable for a polymer melt.
\bibitem{Rg} $R_\mathrm{g}$ is calculated via $<R_\mathrm{g}^2>=N_\mathrm{K}
b_\mathrm{K}^2/\sqrt{6}$, where $b_\mathrm{K}$ denotes the Kuhn
length and $N_\mathrm{K}$ the number of Kuhn monomers. For PS,
$b_\mathrm{K}$=1.8\,nm and
$N_\mathrm{K}$=$M_\mathrm{N}/(720\,$g\,mol$^{-1}$) is found
according to \cite{Rub03}.
\end{thebibliography}
\end{document}